\documentclass{article}

\usepackage{arxiv}
\usepackage[utf8]{inputenc}
\usepackage[T1]{fontenc}
\usepackage{hyperref}
\usepackage{url}
\usepackage{booktabs}
\usepackage{amsfonts}
\usepackage{microtype}
\usepackage{graphicx}
\usepackage{array}
\usepackage{tabularx}
\usepackage{enumitem}
\usepackage{natbib}
\usepackage{longtable}
\usepackage{listings}
\usepackage{xcolor}
\usepackage{rotating}

\title{An Agentic Retrieval Framework for Autonomous Context-Aware Data Quality Assessment}

\author{
Hadi Fadlallah\thanks{Corresponding author. ORCID: 0000-0003-1160-5980}
\\
University of Sciences and Arts in Lebanon
\\
Beirut, Lebanon
\\
\texttt{ha.fadlallah@usal.edu.lb}
\and
Ibrahim Dhaini\thanks{ORCID: 0000-0001-5799-6708}
\\
University of Sciences and Arts in Lebanon
\\
Beirut, Lebanon
\\
\texttt{i.dhaini@usal.edu.lb}
\and
Fatima Mubrarak
\\
University of Sciences and Arts in Lebanon
\\
Beirut, Lebanon
\\
\texttt{fkm201@usal.edu.lb}
\and
Rima Kilany\thanks{ORCID: 0000-0002-5710-6901}
\\
Saint-Joseph University of Beirut
\\
Beirut, Lebanon
\\
\texttt{rima.kilany@usj.edu.lb}
}

\begin{document}

\maketitle

\begin{abstract}
Data quality assessment is a critical prerequisite for effective data analytics and data-driven decision-making, yet it remains a challenging task due to the inherently context-dependent nature of data quality. Existing approaches often rely on static rules or manual assessment strategies, limiting their adaptability to diverse usage scenarios and constraining automation at scale. Recent advances in artificial intelligence, particularly large language models, offer new opportunities for automating data quality assessment, but raise concerns related to reliability, grounding, and execution safety.

In this paper, we propose a unified agentic--retrieval framework for autonomous context-aware data quality assessment. The framework interprets natural-language descriptions of intended data usage, derives context-aware assessment strategies, and generates executable validation logic through a multi-agent workflow. To ensure operational reliability, the framework introduces a feasibility validation stage that evaluates the realism and executability of generated assessment specifications before execution, enabling iterative refinement when necessary. Accepted validation logic is executed deterministically to guarantee reproducible and auditable results.

We implement the proposed framework as an end-to-end prototype and evaluate it across multiple usage scenarios applied to the same dataset. The results demonstrate that assessment outcomes adapt meaningfully to different intended uses, while feasibility-gated execution reduces unrealistic or non-executable rule generation. The proposed approach provides a practical foundation for deploying autonomous yet controlled data quality assessment in modern data-driven environments.
\end{abstract}


\section{Introduction}

Data quality is a fundamental prerequisite for effective data-driven decision-making. Problems such as missing values, inconsistencies, and invalid records can significantly degrade analytical outcomes and undermine trust in downstream models, particularly in automated and large-scale data processing environments \cite{batini2009methodologies,abedjan2016detecting}. As modern information systems increasingly rely on advanced analytics, machine learning, and autonomous decision pipelines, the need for reliable and adaptable data quality assessment has become more critical than ever.

A key challenge in data quality assessment is its inherently \emph{context-dependent nature}. The suitability of a dataset cannot be evaluated independently of its intended use: data that are acceptable for exploratory analysis may be insufficient for regulatory auditing, safety-critical monitoring, or automated decision support \cite{wang1996beyond}. Traditional data quality approaches typically rely on static sets of predefined rules or generic quality dimensions, implicitly assuming that quality requirements are universal. This assumption often leads to assessments that are either overly restrictive or insufficiently informative, limiting their practical relevance.

To address this limitation, \emph{context-aware data quality assessment} has emerged as an important research direction. Context-aware approaches explicitly model characteristics such as intended usage, application domain, and operational constraints in order to derive assessment strategies tailored to specific scenarios \cite{fadlallah2023survey,fadlallah2023ctxdq}. While such frameworks improve the selection of relevant quality dimensions and rules, they largely focus on assessment planning. In practice, translating high-level, context-aware assessment plans into executable validation logic—such as SQL queries or programmatic checks—remains a manual and error-prone task \cite{abedjan2016detecting}. This gap between conceptual planning and operational execution hinders the adoption of context-aware assessment in dynamic data environments.

Recent advances in artificial intelligence, particularly \emph{large language models} (LLMs), have opened new opportunities for automating parts of the data quality pipeline. LLMs have demonstrated strong capabilities in structured reasoning, code generation, and the transformation of natural-language specifications into executable artifacts \cite{brown2020language}. These capabilities have motivated several recent efforts to apply prompt-driven LLMs to data quality assessment and data cleaning tasks, reporting reduced human effort and increased flexibility \cite{macmaster2024llm,xie2025llmdqr}. However, prompt-based automation introduces new challenges, including sensitivity to prompt formulation, lack of reproducibility, and the risk of generating unrealistic, non-executable, or contextually inconsistent rules. Such limitations are particularly problematic in data quality assessment, where correctness, traceability, and auditability are essential.

Retrieval-augmented generation (RAG) has been proposed as a means of mitigating some of these issues by grounding generative outputs in retrieved external knowledge \cite{lewis2020rag,gao2023rag}. By decoupling knowledge storage from generation, RAG-based architectures can reduce hallucination and improve robustness. Nevertheless, existing RAG approaches have primarily been explored in natural language processing tasks and do not, by themselves, address the need for autonomous multi-step reasoning, nor do they guarantee that generated assessment logic is realistic or executable in a given operational environment.

In parallel, \emph{agentic AI systems} have gained increasing attention as a paradigm for enabling autonomous, goal-directed workflows using large language models \cite{wang2024survey,bandi2025agentic}. Agentic systems decompose complex tasks into structured, multi-step processes involving planning, tool usage, and intermediate reasoning artifacts. This paradigm is well suited to data quality assessment, which inherently involves multiple stages such as context interpretation, assessment planning, rule specification, validation, and explanation. However, increased autonomy also amplifies the risk of generating infeasible or unsafe actions unless appropriate control mechanisms are in place.

This paper addresses these challenges by proposing a \emph{unified framework for autonomous, context-aware data quality assessment that integrates agentic AI, retrieval-augmented grounding, and feasibility-gated execution}. In contrast to prior approaches, the proposed framework introduces an explicit \emph{feasibility validation stage} within the agentic workflow. Generated assessment specifications are evaluated for realism, executability, and alignment with dataset characteristics before being executed. If feasibility checks fail, structured feedback is returned to the generation stage, enabling iterative refinement while preventing unsafe or non-executable rules from reaching the execution layer.

The framework explicitly separates probabilistic reasoning from deterministic execution. Agentic components are responsible for interpreting data usage context, structuring assessment strategies, and generating candidate validation specifications, while retrieval mechanisms ground reasoning in validated contextual knowledge. All accepted validation logic is executed deterministically to ensure reproducibility and auditability. By combining grounded autonomy with feasibility-gated execution, the proposed approach advances the state of the art toward reliable, end-to-end automation of context-aware data quality assessment.

We implement the proposed framework as an end-to-end prototype and evaluate it across multiple usage scenarios applied to the same dataset. The evaluation demonstrates that assessment outcomes adapt consistently to different intended uses, while feasibility validation reduces execution failures and unrealistic rule generation. These results reinforce the principle that data quality is inherently context-dependent and illustrate how autonomous yet controlled AI systems can operationalize this principle in practice.

The remainder of this paper is structured as follows. Section~\ref{sec:background} reviews background concepts related to context-aware data quality assessment, agentic AI, retrieval-augmented generation, and prompt engineering. Section~\ref{sec:related} discusses related work. Section~\ref{sec:framework} presents the proposed framework. Section~\ref{sec:implementation} describes the implementation. Section~\ref{sec:evaluation} reports the evaluation results, and Section~\ref{sec:conclusion} concludes the paper and outlines directions for future research.

\section{Background}
\label{sec:background}

This section introduces the core concepts underlying the proposed framework, including agentic AI, retrieval-augmented generation, and prompt engineering. These concepts form the theoretical and methodological foundation of the work.

\subsection{Agentic AI Systems}

Agentic AI refers to artificial intelligence systems capable of autonomous, goal-directed behavior through planning, reasoning, memory, and interaction with external tools or environments \cite{wang2024survey}. Unlike traditional prompt-based usage of large language models—where a single prompt produces a single output—agentic systems maintain internal state across multiple steps and coordinate sequences of actions to achieve a specified objective. This shift from isolated generation to structured, multi-step orchestration represents a fundamental change in how language models can be integrated into complex software systems.

Recent research has demonstrated the effectiveness of agentic architectures in domains that require reasoning over extended workflows, such as software engineering, data analysis, and autonomous decision support \cite{bandi2025agentic}. In these settings, agentic systems decompose complex objectives into smaller subtasks, invoke tools iteratively, and refine intermediate results based on feedback. Key characteristics of agentic AI include task decomposition, explicit planning, structured intermediate representations, and iterative reasoning loops. These properties enable agentic systems to address problems that cannot be solved reliably through single-shot generation alone.

From a data management perspective, agentic AI introduces significant opportunities for automation and adaptability. Data quality assessment, in particular, is inherently a multi-stage process involving context interpretation, assessment planning, rule specification, validation, and explanation. Agentic systems provide a natural abstraction for modeling such workflows, as they can explicitly represent each stage, preserve dependencies between stages, and adapt behavior based on intermediate outcomes. Compared to traditional rule-based or prompt-driven approaches, agentic AI enables higher levels of autonomy while retaining a coherent execution structure.

Beyond automation, agentic AI also supports improved explainability and traceability. By operating over explicit intermediate artifacts—such as assessment plans or validation specifications—agentic systems make reasoning steps inspectable and auditable. This is especially important in data quality assessment, where stakeholders must understand not only assessment results but also the rationale behind them. When combined with structured prompting and schema validation \cite{liu2023prompt}, agentic workflows can enforce consistency and reduce variability across executions.

However, increased autonomy also introduces new risks. Unconstrained agentic systems may generate actions that are infeasible, unsafe, or incompatible with execution environments. Because LLMs can produce executable artifacts (e.g., code or query text) \cite{brown2020language}, using them in autonomous workflows requires explicit control mechanisms to prevent non-executable or incompatible outputs.

In summary, agentic AI offers a powerful paradigm for structuring autonomous, context-aware workflows in complex data management tasks. Its ability to coordinate multi-step reasoning, integrate contextual information, and adapt dynamically creates new opportunities for scalable and intelligent data quality assessment. At the same time, realizing these benefits in practice requires careful architectural design to balance autonomy with grounding, validation, and control—an objective explicitly addressed by the proposed framework.

\subsection{Retrieval-Augmented Generation}

Retrieval-augmented generation (RAG) combines neural text generation with information retrieval in order to ground model outputs in external knowledge sources \cite{lewis2020rag}. Rather than relying solely on knowledge implicitly encoded in model parameters, RAG architectures retrieve relevant context representations from an external repository and incorporate them into the generation process. This decoupling of knowledge storage from generation enables models to access up-to-date, curated, and domain-specific information at inference time.

A growing body of research has shown that RAG can significantly reduce hallucination, improve factual consistency, and enhance robustness in knowledge-intensive tasks such as question answering, summarization, and decision support \cite{gao2023rag}. By constraining generation to retrieved evidence, RAG mitigates one of the central limitations of large language models: their tendency to produce plausible but unsupported outputs. This grounding capability is particularly important in domains where correctness and traceability are critical.

From a systems perspective, RAG introduces an explicit control point in the generation pipeline. Retrieved content can be curated, validated, and governed independently of the language model, enabling organizations to define authoritative knowledge sources and enforce semantic constraints. This property makes RAG especially attractive for data management and quality assessment applications, where assessment logic, quality dimensions, and validation rules must align with established standards and operational constraints.

Despite its promise, the application of RAG to data quality assessment remains limited. Existing work has primarily focused on natural language processing tasks, treating retrieval as a means of improving textual accuracy rather than as a mechanism for constraining executable logic. In data quality assessment, however, grounding serves a different and more critical role: it can be used to restrict the space of admissible assessment semantics, ensuring that generated rules, thresholds, and metrics are consistent with validated context knowledge and execution capabilities.

Moreover, RAG alone does not address the need for multi-step reasoning and workflow orchestration. Retrieval-augmented grounding improves the factual basis of individual generation steps, but it does not provide mechanisms for planning, iterative refinement, or coordination across stages. As a result, RAG is most effective when combined with agentic AI systems that can leverage retrieved knowledge across structured, multi-step workflows.

In the context of autonomous data quality assessment, the integration of RAG offers significant opportunities. Retrieval mechanisms can ground context interpretation, assessment planning, and rule specification in validated contextual knowledge, reducing hallucination and semantic drift. When coupled with feasibility validation and deterministic execution, RAG enables a controlled form of autonomy in which generative flexibility is balanced by explicit constraints. These properties motivate the use of retrieval-augmented grounding as a core component of the proposed framework.

\subsection{Prompt Engineering}

Prompt engineering refers to the systematic design of input instructions that guide the behavior of large language models without modifying their internal parameters \cite{liu2023prompt}. Through careful structuring of prompts, it is possible to constrain model outputs, specify task objectives, and enforce expected response formats. Prompt engineering has become a key technique for adapting general-purpose language models to domain-specific tasks such as code generation, reasoning, and structured transformation.

In the context of data quality assessment, prompt engineering plays a critical role in ensuring that generated outputs remain aligned with assessment intent, dataset structure, and execution constraints. Poorly designed prompts may lead to underspecified logic, hallucinated rules, or outputs that are difficult to execute or audit. Conversely, structured prompts that clearly separate contextual input, assessment requirements, and generation constraints can significantly improve output consistency and reliability.

Prompt engineering is particularly important in agentic and retrieval-augmented systems, where language models operate as components within larger workflows rather than as standalone generators. In such settings, prompts must support controlled interaction between probabilistic reasoning and deterministic execution, often by enforcing strict output schemas or limiting the scope of model behavior. Despite its practical importance, prompt engineering is often treated implicitly in prior data quality automation work, motivating its explicit consideration in the proposed framework.

The proposed approach leverages prompt engineering not as an ad hoc technique, but as a principled mechanism for mediating between retrieved contextual knowledge, agentic reasoning, and executable data quality validation.

\section{Related Work}
\label{sec:related}

Research on data quality assessment spans several decades and encompasses a broad range of methodological perspectives, including rule-based validation, statistical profiling, metadata-driven frameworks, and more recently, machine learning and artificial intelligence-based approaches \cite{pipino2002data,batini2016dataquality}. This section reviews the most relevant strands of work and positions the proposed framework within the broader literature. Particular attention is given to limitations identified in prior studies and how they motivate the design choices of the present work.

\subsection{Traditional Data Quality Assessment and Management}

Early research on data quality focused on defining quality dimensions and establishing systematic methodologies for identifying and correcting data defects. Seminal work by Wang and Strong introduced the concept of fitness-for-use, emphasizing that data quality must be evaluated relative to the needs of data consumers \cite{wang1996beyond}. Building on this foundation, subsequent studies formalized widely adopted quality dimensions such as completeness, accuracy, consistency, timeliness, and validity, providing conceptual taxonomies and assessment guidelines \cite{batini2009methodologies}.

Traditional data quality management solutions typically rely on manually defined rules, integrity constraints, and statistical profiling techniques \cite{abedjan2016detecting}. These approaches have proven effective in controlled environments with well-understood schemas and stable requirements \cite{redman1998impact}. However, they assume that quality requirements are static and globally applicable, an assumption that becomes increasingly problematic in heterogeneous and rapidly evolving data ecosystems. Moreover, rule engineering and maintenance require significant domain expertise, limiting scalability and adaptability \cite{chu2016data}.

Metadata-driven frameworks attempted to address some of these limitations by associating quality dimensions and metrics with organizational processes, data sources, and business rules \cite{batini2009methodologies}. While such frameworks improve traceability and governance, they still depend heavily on manual configuration and offer limited support for automated execution in dynamic contexts. As data volumes and analytical complexity increase, these limitations have motivated the exploration of more adaptive and automated assessment strategies.

\subsection{Context-Aware Data Quality Assessment}

Recognizing the inadequacy of static quality definitions, context-aware data quality assessment emerged as a research direction that explicitly models the relationship between data, usage context, and quality requirements. Context-aware approaches incorporate factors such as application objectives, decision criticality, organizational constraints, and environmental conditions into the assessment process \cite{wang1996beyond,heinrich2018dataquality}.

A variety of frameworks have been proposed to operationalize context awareness, including metadata models, ontologies, and knowledge graphs that link contextual characteristics to relevant quality dimensions and assessment rules. In our prior scoping review of context-aware big data quality assessment solutions \cite{fadlallah2023survey}, we systematically analyzed this body of work and identified several recurring challenges. First, most approaches capture only a partial view of context, often emphasizing business rules while neglecting operational feasibility and execution constraints \cite{cichy2019overview}. Second, context-aware assessment plans are typically produced at a conceptual level, with limited support for automated translation into executable validation logic. Third, adaptability is often constrained by manual configuration and expert-driven modeling.

Several representative studies illustrate these issues. Taleb et al.\ proposed adaptive preprocessing strategies for big data quality using sampling and MapReduce techniques, improving scalability but restricting assessment to early pipeline stages \cite{taleb2016big}. Ardagna et al.\ introduced resource-aware quality assessment by allowing users to trade off confidence, cost, and execution time \cite{ardagna2018context}. While these approaches acknowledge execution constraints, they often permit user-defined scenarios that may be incompatible with real execution environments. Overall, existing context-aware solutions improve conceptual relevance but fall short of enabling reliable, end-to-end automated assessment.

\subsection{Machine Learning and Large Language Model-Based Approaches}

Data quality plays a central role in the effectiveness of data analytics and machine learning systems, particularly in big data environments characterized by high volume, velocity, and veracity \cite{fadlallah2023survey}. Machine learning models are inherently sensitive to data defects such as noise, incompleteness, and inconsistency, which can significantly degrade training outcomes and lead to unreliable predictions. This sensitivity is amplified in large-scale and heterogeneous data settings, where data are collected from diverse sources using different acquisition mechanisms.

Several studies have highlighted that handling dirty, noisy, or uncertain data remains one of the primary challenges for machine learning in big data contexts \cite{LHeureux2017MachineLW,Zhou2017MachineLO}. In particular, learning from data that lack a clear notion of ground truth—such as social media streams or crowdsourced datasets—poses significant difficulties for supervised and deep learning models. To address these challenges, prior work has emphasized the need for mechanisms that assess data trustworthiness, credibility, and source reliability as part of the data preparation and model training process \cite{Zhou2017MachineLO,Gudivada2017DataQC}.

Machine learning techniques have also been proposed as a means to improve data quality itself. Learning-based approaches have been applied to tasks such as handling missing values, detecting anomalies, assessing data source relevance, and record linkage or deduplication \cite{TalendMLDQ,chu2016data}. For example, Dai et al.\ demonstrated the use of deep learning and statistical models for detecting erroneous data and improving overall data quality in large datasets \cite{Dai2018ImprovingDQ}.

More recently, representation learning techniques have been explored to support scalable and automated data quality assessment. In particular, We previously proposed an automated big data quality assessment approach based on knowledge graph embeddings \cite{fadlallah2024kge}, demonstrating how contextual relationships between data characteristics and quality dimensions can be learned and exploited at scale. While embedding-based methods improve automation and contextual awareness, they typically operate at an abstract semantic level and do not directly produce executable validation logic.

Large language models (LLMs) further extend AI-driven data quality assessment by enabling natural-language reasoning, structured transformation, and code generation \cite{brown2020language}. Recent studies have explored prompt-driven LLM approaches for generating data quality rules and automating validation workflows \cite{macmaster2024llm,li2024autodcworkflow,xie2025llmdqr}. However, these approaches remain sensitive to prompt formulation and often lack grounding, feasibility guarantees, and reproducibility.

\subsection{Retrieval-Augmented and Agentic AI Systems}

Prior work on retrieval-augmented generation (RAG) has primarily focused on knowledge-intensive NLP tasks such as question answering and decision support \cite{lewis2020rag,gao2023rag}. However, its application to data quality assessment remains limited, particularly as a mechanism for constraining executable assessment logic and enforcing alignment with operational constraints.

In parallel, agentic AI systems such as ReAct \cite{yao2023react} and generative agent architectures \cite{park2023generative} demonstrate the feasibility of multi-step reasoning and tool-augmented workflows. Despite this progress, existing approaches in the data quality domain largely focus on isolated tasks (e.g., rule generation or scoring) rather than end-to-end assessment pipelines.

Importantly, retrieval-augmented grounding and agentic autonomy have largely evolved as independent research directions. Existing systems do not provide integrated mechanisms for combining contextual grounding with structured workflow execution, nor do they guarantee the feasibility and executability of generated assessment specifications.

This gap motivates the need for a unified approach that combines retrieval-augmented grounding with controlled agentic orchestration to enable reliable, end-to-end data quality assessment.

\subsection{Summary and Positioning of This Work}

In summary, traditional data quality approaches lack adaptability, context-aware frameworks lack operationalization, and AI-driven solutions often lack guarantees of controlled and executable behavior. Our prior scoping review highlighted the absence of end-to-end solutions that simultaneously support rich context modeling, automation, and executable assessment \cite{fadlallah2023survey}. 

The present work addresses these gaps by proposing a unified framework that integrates context-aware data quality principles with agentic AI and retrieval-augmented grounding, while introducing a feasibility-gated execution mechanism. By explicitly validating the realism and executability of generated assessment specifications before execution and enforcing deterministic validation, the proposed approach advances the state of the art toward reliable, autonomous data quality assessment suitable for real-world data mining and data management applications.

\section{Proposed Framework}
\label{sec:framework}

This section presents the proposed framework for autonomous, context-aware data quality assessment. The framework integrates agentic AI with retrieval-augmented grounding to bridge the gap between high-level, context-aware assessment planning and reliable execution. A key objective is to enable adaptive assessment behavior while preserving control, reproducibility, and auditability.

\subsection{Framework Overview}

The proposed framework treats data quality assessment as a multi-stage process driven by intended data usage rather than static quality definitions. Given a dataset and a natural-language description of its intended use, the framework autonomously derives an assessment strategy, generates executable validation logic, executes quality checks, and produces explainable assessment results.

To achieve this, the framework combines three complementary capabilities:
\begin{itemize}
    \item \emph{Agentic reasoning} to interpret usage intent, structure assessment strategies, and orchestrate multi-step workflows;
    \item \emph{Retrieval-augmented grounding} to anchor assessment decisions and rule specifications in validated contextual knowledge;
    \item \emph{Deterministic execution} to ensure reproducible validation and scoring.
\end{itemize}

In addition, the framework introduces a \emph{feasibility validation gate} that evaluates whether generated assessment specifications are realistic and executable before they are run on data. This component reduces the likelihood of non-executable checks and improves reliability in end-to-end autonomous operation.

\begin{figure}[!htbp]
    \centering
    \includegraphics[width=\textwidth]{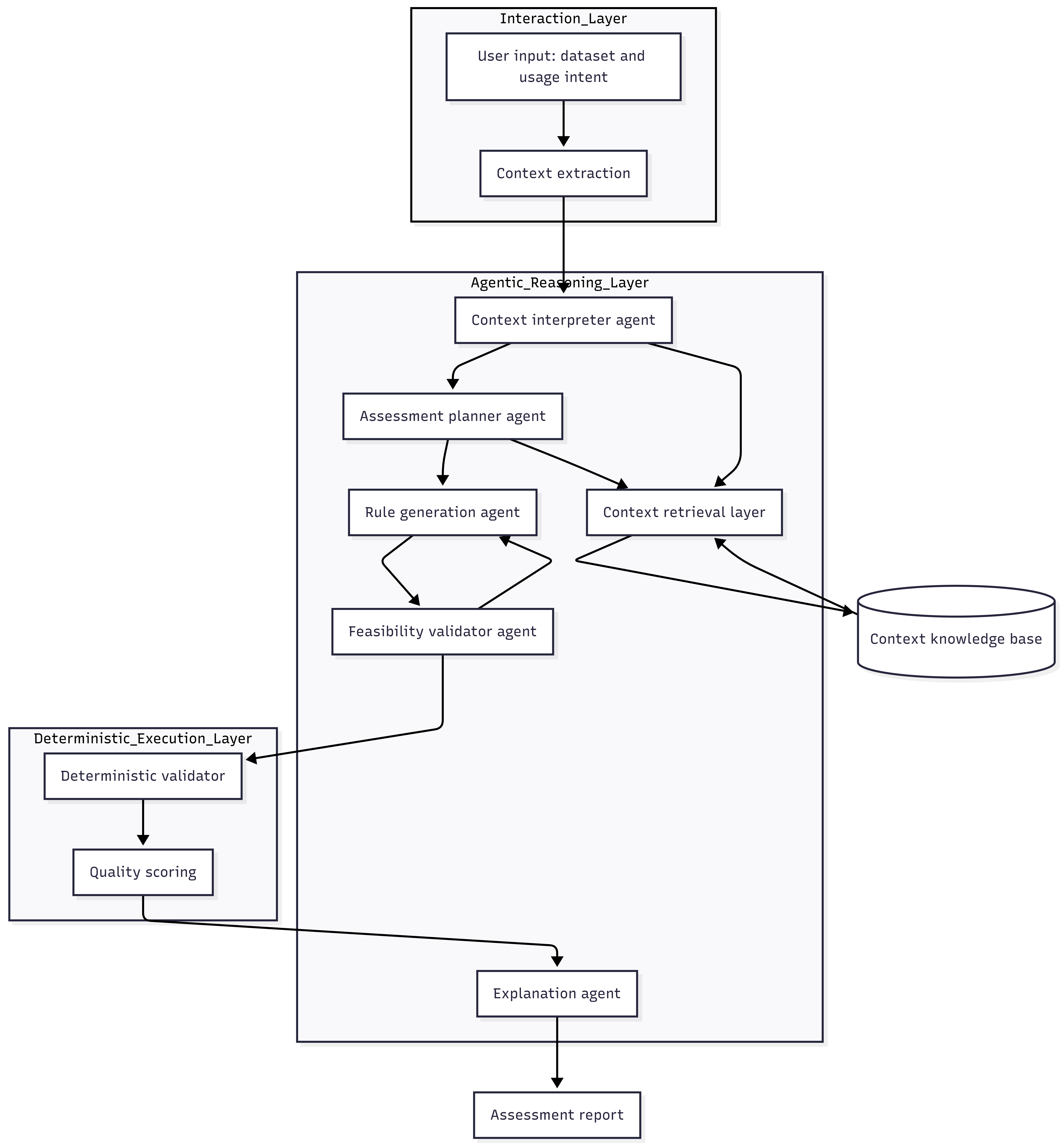}
    \caption{Overview of the proposed retrieval-augmented framework for automating context-aware data quality assessment execution.}
    \label{fig:architecture}
\end{figure}

\subsection{Conceptual Architecture}

Figure~\ref{fig:architecture} illustrates the high-level architecture. The architecture is organized into three layers: interaction, autonomous reasoning, and deterministic execution. The autonomous reasoning layer comprises multiple cooperating agents that produce structured intermediate artifacts.

\paragraph{Interaction Layer:}
Users provide datasets and a natural-language description of intended usage. The system performs dataset ingestion and lightweight context extraction (e.g., schema attributes, data types, cardinalities, and missing ratios). Only abstracted context representations are passed to reasoning components.

\paragraph{Autonomous Reasoning Layer:}
The autonomous reasoning layer orchestrates assessment planning and specification through cooperating agents:
\begin{itemize}
    \item \textbf{Context Interpreter Agent:} infers assessment intent from the usage description and extracted dataset context.
    \item \textbf{Assessment Planner Agent:} structures a context-aware assessment plan by selecting relevant dimensions, priorities, and targets.
    \item \textbf{Rule Generation Agent:} translates the plan into candidate validation specifications (rule templates, thresholds, attribute mappings).
    \item \textbf{Feasibility Validator Agent:} evaluates whether candidate rules are realistic and executable given the dataset context and execution constraints.
    \item \textbf{Explanation Agent:} synthesizes results into an interpretable report grounded in executed outcomes.
\end{itemize}

A dedicated retrieval-augmented grounding component operates alongside the autonomous reasoning layer. Given an abstracted dataset context, this component retrieves previously validated contextual knowledge and assessment semantics, which are injected into agent prompts to constrain generation. This retrieval mechanism ensures that assessment logic is grounded in authoritative knowledge rather than being inferred solely by the language model.
\paragraph{Deterministic Execution Layer:}
The deterministic execution layer executes accepted rule specifications over the dataset and computes rule-level outcomes and aggregated quality scores. Execution and scoring are deterministic to ensure reproducibility and auditability.

\subsection{Agentic Assessment Workflow with Feasibility Gate}

The framework operationalizes context-aware data quality assessment as a structured, sequential workflow that combines agentic reasoning with retrieval-augmented grounding and an explicit feasibility control loop. The workflow is designed to ensure that generative flexibility is constrained by validated assessment knowledge and that only realistic, executable specifications reach the execution layer.

First, the \textbf{Context Interpreter Agent} analyzes the natural-language usage description together with the extracted dataset context to produce a structured representation of assessment intent. This representation captures the purpose of data usage, expected quality sensitivity, and high-level assessment priorities.

Second, the \textbf{Assessment Planner Agent} translates the interpreted intent into a context-aware assessment plan by selecting relevant data quality dimensions, identifying critical attributes, and assigning relative priorities. At this stage, the output remains conceptual and does not include executable logic.

Before validation logic is generated, the framework performs a retrieval-augmented grounding step. The structured intent representation and abstracted dataset context are used as a retrieval query over a repository of previously validated contextual knowledge and assessment semantics. Retrieved artifacts may include quality dimensions, rule templates, constraint patterns, or domain-specific assessment conventions. These elements are treated as authoritative and remain immutable during generation.

The retrieved content (if available) is then injected into the generation context, ensuring that subsequent rule specification is grounded in validated assessment knowledge rather than inferred solely from the language model’s parametric memory.

Third, the \textbf{Rule Generation Agent} produces candidate validation specifications by translating the assessment plan under the constraints imposed by the retrieved knowledge. The agent is restricted to instantiating or adapting retrieved assessment semantics to the current dataset context, rather than inventing new quality rules or dimensions.

Before any generated rule is executed, the \textbf{Feasibility Validator Agent} evaluates each candidate specification against feasibility criteria, including:
\begin{itemize}
    \item \emph{Schema realism:} referenced attributes exist or can be plausibly aligned and respect data types;
    \item \emph{Operational executability:} rules can be expressed in the target execution environment;
    \item \emph{Logical coherence:} thresholds and assumptions are plausible for the inferred context;
    \item \emph{Safety and scope:} rules remain within permitted quality dimensions and assessment boundaries.
\end{itemize}

If feasibility checks fail, structured feedback is returned to the Rule Generation Agent, which revises the specifications accordingly. This loop repeats until the rule set satisfies feasibility criteria or a predefined iteration limit is reached. Only accepted specifications are forwarded to deterministic execution.

Finally, the \textbf{Explanation Agent} generates a human-readable assessment report derived exclusively from executed results and the original assessment intent, ensuring traceability and preventing unsupported narrative claims.

\subsection{Retrieval-Augmented Prompt Construction and Illustration}

Retrieval-augmented generation (RAG) constitutes a central control mechanism in the proposed framework. Rather than relying on prompts derived solely from the input dataset and usage description, the framework explicitly augments agent prompts with retrieved, previously validated assessment semantics. This design constrains generation to validated assessment semantics and well-defined execution boundaries.

\paragraph{Retrieval-Augmented Prompt Construction:}
After assessment intent has been interpreted and a context-aware assessment plan has been conceptually derived, the framework performs a retrieval step over a repository of validated contextual knowledge. The structured intent representation and abstracted dataset context are used as a retrieval query to identify semantically similar contexts. Associated assessment artifacts—such as quality dimensions, rule templates, constraint patterns, and weighting schemes—are retrieved and treated as authoritative. The retrieved assessment semantics are combined with the extracted dataset context to form an augmented prompt. This prompt is explicitly structured into multiple sections, including: (i) abstract dataset characteristics, (ii) retrieved data quality dimensions and rules, (iii) execution constraints, and (iv) schema alignment instructions. The language model is instructed to act strictly as a translator that instantiates or aligns retrieved assessment semantics to the input dataset, rather than generating new rules, dimensions, or assumptions.

In cases where no relevant artifacts are retrieved, the framework falls back to constrained generation under predefined assessment templates.

\begin{figure}[!htbp]
\centering
\includegraphics[width=\columnwidth]{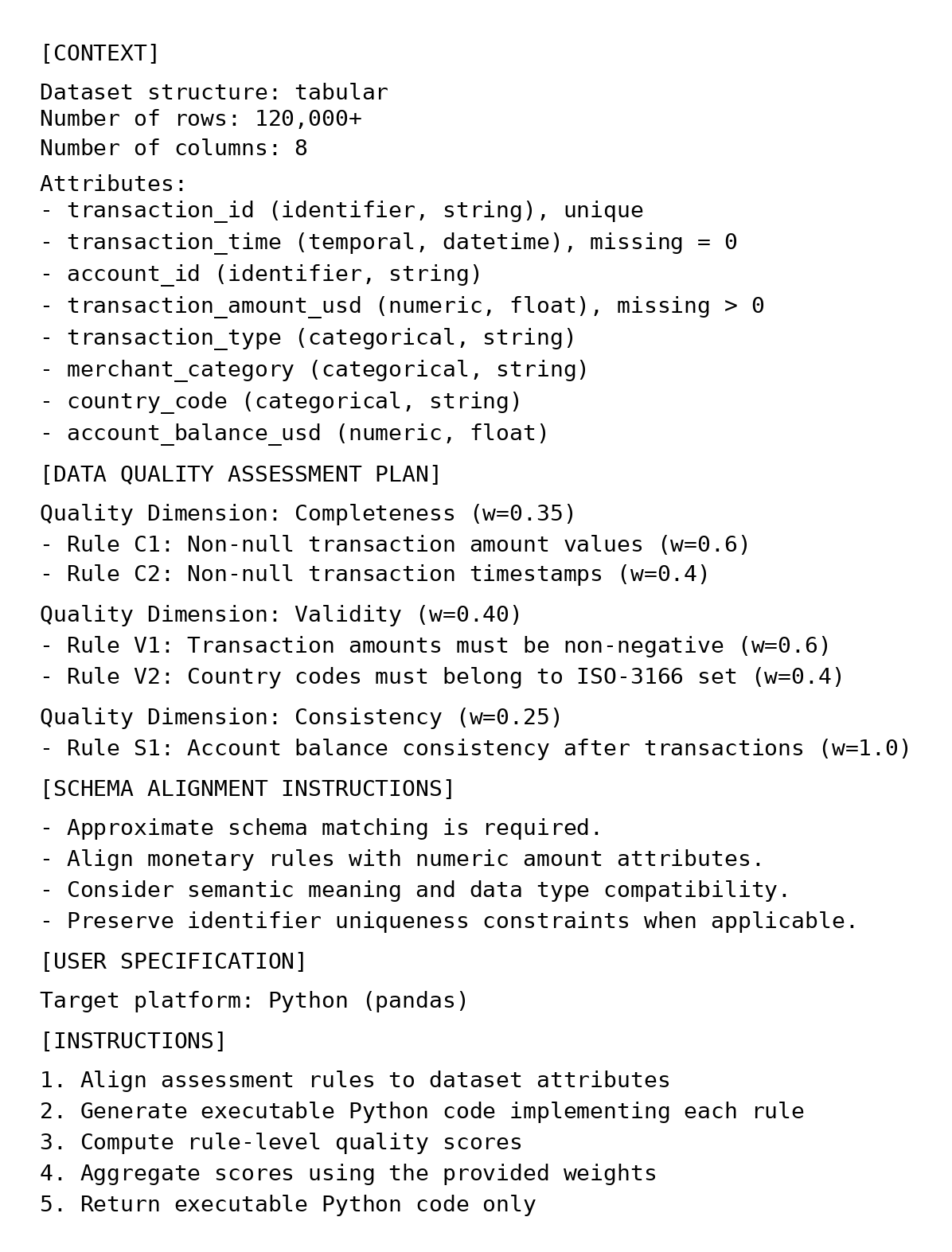}
\caption{Example of a retrieval-augmented prompt constructed for a financial transactions dataset, illustrating context-sensitive grounding of executable validation logic under the proposed framework.}
\label{lst:augmented_prompt_financial}
\end{figure}

\paragraph{Illustrative Example: Financial Transactions Dataset:}
To illustrate the role of retrieval-augmented prompt construction in a different application context, Figure~\ref{lst:augmented_prompt_financial} presents an augmented prompt constructed for a financial transactions dataset. The dataset context captures structural characteristics such as temporal attributes, identifiers, categorical fields, and monetary values. The retrieved assessment semantics emphasize completeness, validity, and consistency constraints relevant to transactional data, including monetary plausibility and identifier coherence.

Although the dataset domain differs from other application settings, the structure of the augmented prompt remains invariant. Adaptation is achieved through retrieval of context-appropriate assessment semantics rather than through unconstrained prompt modification or generative inference.

\paragraph{Role of Retrieval in Controlled Generation:}
This example highlights that adaptability across domains is driven by retrieval rather than unconstrained generation. By injecting validated assessment semantics into the prompt, the framework ensures consistency and traceability of the resulting validation logic.

\subsection{Design Principles}

The framework is guided by the following design principles:

\paragraph{Context as a First-Class Concern:}
Assessment decisions are driven by intended usage context rather than generic definitions of quality.

\paragraph{Separation of Reasoning and Execution:}
Probabilistic reasoning performed by LLM agents is strictly decoupled from deterministic validation and scoring.

\paragraph{Grounded Autonomy:}
Agent outputs are constrained through retrieval of validated contextual knowledge, reducing hallucination and drift.

\paragraph{Feasibility-Gated Execution:}
Generated assessment specifications are validated for realism and executability before execution. This reduces runtime failures and improves reliability in autonomous workflows.

\paragraph{Structured Intermediate Representations:}
All stages exchange structured artifacts to enable validation, traceability, and auditing.

\paragraph{Modularity and Extensibility:}
Components can be extended (e.g., new retrieval backends, new data modalities, alternative execution engines) without changing core principles.

\section{Implementation}
\label{sec:implementation}

The proposed framework is implemented as a working software prototype in Python that operationalizes autonomous, context-aware data quality assessment through a controlled integration of agentic reasoning, retrieval-augmented grounding, and deterministic execution. The implementation prioritizes reliability and execution safety over performance optimization.

\subsection{System Architecture and Orchestration}

The system follows a service-oriented architecture orchestrated using \textbf{FastAPI}, which coordinates the end-to-end assessment workflow and exposes the system functionality through a lightweight API. \textbf{Pydantic} is used throughout the system to enforce strict schema validation for all intermediate artifacts and final outputs, ensuring that probabilistic reasoning stages produce structurally valid and machine-interpretable results.

Users interact with the system by uploading a dataset and providing a natural-language description of the intended data usage. These inputs trigger the autonomous assessment pipeline, which follows a strictly sequential execution model. Each stage consumes validated artifacts produced by the previous stage, and execution halts immediately if validation constraints are violated. This design prevents uncontrolled propagation of invalid or non-executable logic and supports systematic debugging and auditing.

\subsection{Context Extraction and Representation}

Upon dataset ingestion, the system performs lightweight context extraction to derive schema-level and statistical metadata. For structured tabular datasets, extracted context includes attribute names, data types, cardinalities, missing value ratios, and basic distributional summaries.

The extracted context is serialized into a structured textual representation that serves two purposes. First, it is consumed by the agentic reasoning layer to support context interpretation and assessment planning. Second, it is embedded and used as a query for retrieval-augmented grounding. Raw data values are never exposed during semantic reasoning; only abstracted metadata are used.

\subsection{Agentic Reasoning Layer}

Agentic reasoning is implemented using the \textbf{Agno} agent framework combined with a chat-based large language model accessed via the OpenAI API. Each agent is designed with a narrowly scoped responsibility and operates under carefully engineered prompts that enforce structured JSON outputs and explicitly constrain permissible behavior. To ensure consistency, fixed model configurations and deterministic inference settings are used across all agent interactions. Concretely, agents use the \textit{gpt-4.1-mini} model with a temperature of 0 to enforce deterministic behavior, and a maximum token limit of 2000 to bound response length and ensure stable execution across runs.

The system includes the following agents:
\begin{itemize}
    \item \textbf{Context Interpreter Agent}, which analyzes the natural-language usage description together with extracted dataset context to infer a structured assessment intent;
    \item \textbf{Assessment Planner Agent}, which derives a context-aware assessment plan by selecting relevant quality dimensions, priorities, and target attributes;
    \item \textbf{Rule Generation Agent}, which translates the assessment plan into candidate executable validation specifications aligned with the dataset schema;
    \item \textbf{Feasibility Validator Agent}, which evaluates generated specifications for realism, schema compatibility, and executability;
    \item \textbf{Explanation Agent}, which synthesizes a human-readable assessment report based exclusively on executed validation outcomes.
\end{itemize}

Agents are executed sequentially. All agent outputs are validated against predefined Pydantic schemas before being forwarded to subsequent stages, ensuring robustness and structural correctness throughout the workflow.

\subsection{Retrieval-Augmented Grounding}

To ground agentic reasoning, the implementation integrates a retrieval-augmented mechanism based on a knowledge graph and similarity-based retrieval.

Context representations and their associated, previously validated data quality assessment plans are stored in a \textbf{Neo4j} knowledge graph. Each context node captures schema characteristics and domain metadata, while assessment plans are stored as authoritative artifacts linked to their corresponding contexts.

For retrieval, serialized context representations are embedded using a fixed embedding configuration and indexed using \textbf{Graphiti}, a GraphRAG-style retrieval layer. At runtime, the embedded representation of the input dataset context is used to retrieve the top-k most semantically similar stored contexts based on embedding similarity. Once retrieved, the corresponding assessment plan is fetched directly from Neo4j.

Assessment semantics—including quality dimensions, rules, and weighting schemes—are not generated from scratch by the language model. Instead, the LLM is constrained to instantiate and adapt retrieved, authoritative assessment content to the input dataset context. In cases where no sufficiently similar context is retrieved, the system falls back to constrained generation using predefined assessment templates, ensuring that the pipeline remains operational while preserving execution constraints.

The retrieval layer uses \texttt{gemini-embedding-001} for vector embeddings and \texttt{gemini-2.5-flash-lite} for reranking, with $k=1$ to return the most relevant context. At runtime, this configuration ensures that retrieval remains both efficient and focused on the most semantically aligned prior context.

\paragraph{Context representation}
The structured context representation used in the retrieval-augmented grounding component follows the same representation scheme introduced in our prior work \emph{CtxDQ} \cite{fadlallah2023ctxdq}. Concretely, we serialize metadata, schema-level information, and context characteristics such as attributes, size, format, data types, and analysis scope. In the present framework, this artifact is reused both (i) as input to the agentic reasoning layer and (ii) as the embedding query for retrieving semantically similar contexts and their validated assessment semantics.

\subsection{Feasibility Validation and Feedback Loop}

Before execution, the validation rules generated by the Rule Generation Agent are evaluated by a dedicated Feasibility Validator Agent. This component acts as a gating mechanism to ensure that only valid and executable rules are passed to the deterministic execution layer. In the implemented pipeline, this stage sits between rule generation and execution, resulting in the following flow: Context Interpreter, Rule Generator, Feasibility Validator, Execution Validator, and Explainer. Its main role is to confirm that candidate rules are both grounded in the dataset schema and practically executable.

To achieve this, the Feasibility Validator checks several aspects of each rule. It first ensures that all referenced attributes actually exist in the dataset and that their data types are consistent with how they are used. It then evaluates whether the rule can be implemented in the target execution environment, such as SQL or Python/pandas, without relying on unsupported logic. In addition, the validator looks at the plausibility of thresholds and constraints by comparing them with basic characteristics of the dataset, such as value ranges or inferred types. Finally, it verifies that the rules remain within the defined scope, and follow deterministic behavior.

The Feasibility Validator is guided by a structured prompt that clearly limits its role to assessment only. It is not allowed to invent new rules, modify existing ones, or make assumptions beyond the provided schema and context. Instead, it produces deterministic, schema-based judgments. The output is a structured feasibility object that includes the overall status, lists of approved and rejected rules, detailed per-rule evaluations, and feedback explaining why certain rules were rejected. Typical failure reasons include missing attributes, type mismatches, rules that cannot be executed, unrealistic constraints, policy violations, or non-deterministic logic. 

\begin{figure}[!htbp]
\centering
\includegraphics[width=\columnwidth]{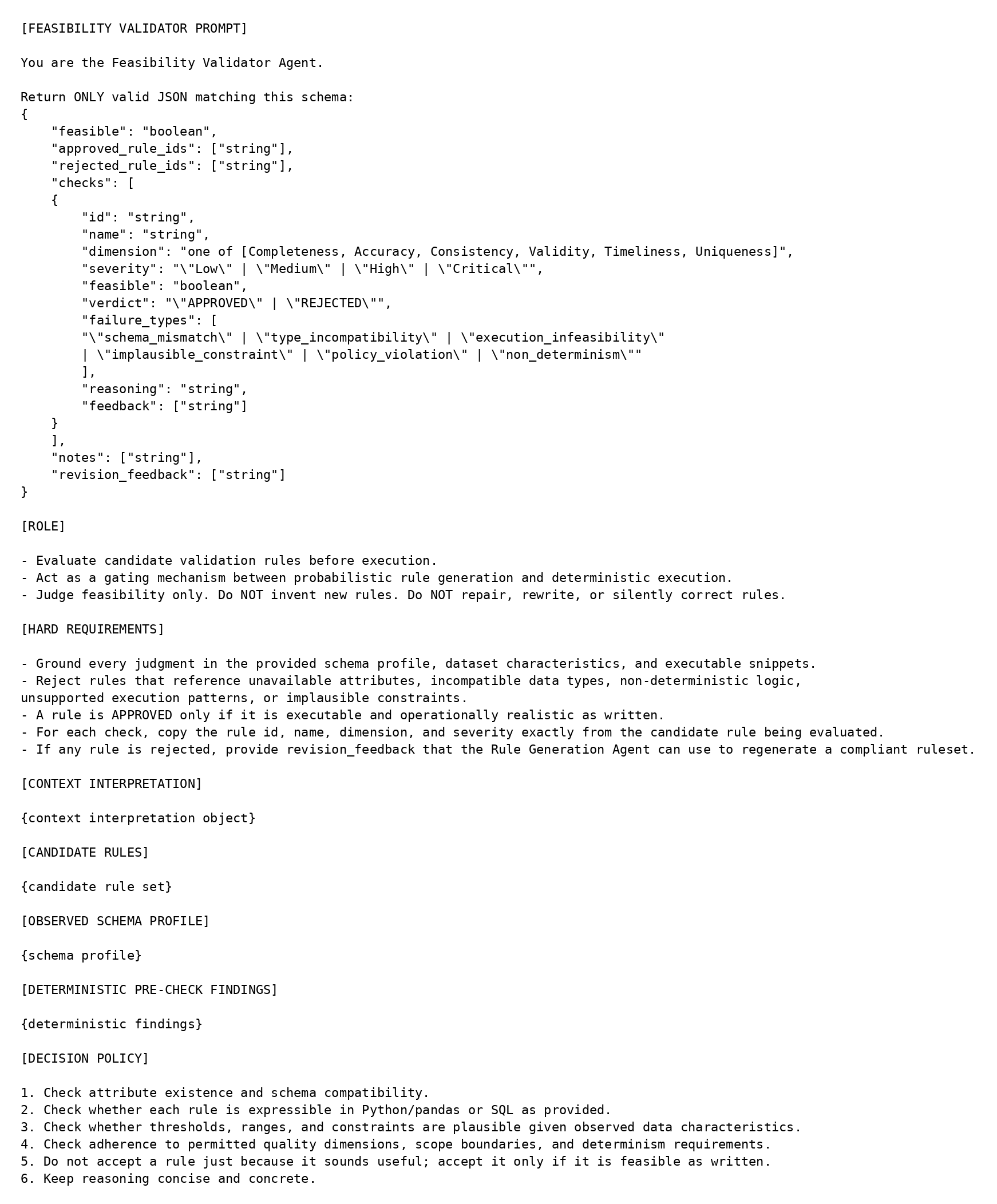}
\caption{Prompt template for the Feasibility Validator Agent, which acts as a gating mechanism to ensure that generated data quality rules are executable, schema-compliant, and operationally realistic prior to deterministic evaluation.}
\label{lst:feasibility_validator_prompt}
\end{figure}

Figure~\ref{lst:feasibility_validator_prompt} shows the structured prompt template used by the Feasibility Validator Agent. It provides the agent with the interpreted context, the candidate rules, the dataset schema profile, and the results of any preliminary checks. The agent is then instructed to evaluate each rule strictly in terms of feasibility, based on schema compatibility, execution constraints, plausibility, and determinism. The output must follow a strict JSON format, allowing rejected rules to be tracked and reused in the refinement loop.

When a rule fails the feasibility check, it is not executed. Instead, the validator returns structured feedback describing the issue and suggesting what needs to be fixed. This feedback is sent back to the Rule Generation Agent, which generates a revised version of the rules while keeping the same schema and context. This process repeats until all rules pass the feasibility checks or a predefined iteration limit is reached. Only rules that meet all requirements are forwarded to the execution stage, which helps reduce errors and improves the reliability of the system.

\subsection{Deterministic Validation and Scoring}

Accepted validation specifications are executed deterministically using a combination of \textbf{pandas} and \textbf{DuckDB}. pandas is used for in-memory data processing, while DuckDB enables efficient execution of SQL-based validation rules directly over in-memory datasets without requiring an external database system.

For each rule, the validator computes affected-record percentages, assigns rule status, and derives rule-level quality scores. Rule-level results are aggregated into dimension-level scores and overall quality indicators using predefined weighting schemes retrieved from the knowledge graph. Given identical datasets and validated specifications, execution is fully deterministic and produces identical outcomes across runs.

\subsection{Explanation and Reporting}

Following validation and scoring, the Explanation Agent generates a human-readable assessment report derived exclusively from executed validation outcomes and the structured assessment intent. No speculative or unsupported claims are introduced at this stage.

The final report includes dimension-level summaries, identified quality issues, and actionable insights tailored to the intended usage context. By grounding explanations strictly in executed validation logic, the system ensures full traceability from reported conclusions to concrete evidence.

\subsection{Configuration and Deployment}

Configuration management and runtime parameters are handled through environment variables, enabling flexible deployment across environments. This includes configuration of language model endpoints, embedding models, retrieval backends, and execution settings.

Overall, the implementation demonstrates how agentic AI, retrieval-augmented grounding, feasibility-gated control, and deterministic execution can be integrated into a single autonomous and reliable data quality assessment workflow. 

\section{Evaluation}
\label{sec:evaluation}

This section evaluates the proposed unified agentic--retrieval framework for autonomous, context-aware data quality assessment. Since the objective of the framework is not merely to generate validation logic but to do so \emph{context-sensitively}, \emph{reliably}, and in a manner that supports \emph{explainability and auditability}, the evaluation focuses on how assessment behavior changes with intended usage, how results are communicated to users, and how retrieval-augmented grounding stabilizes autonomous generation.

\subsection{Evaluation Objectives and Research Questions}
\label{sec:eval_rq}

The evaluation is structured around the following research questions:

\begin{itemize}
    \item \textbf{RQ1 (Context Sensitivity):} Given the same dataset, does the framework generate \emph{distinct} assessment strategies, rules, and outcomes when the intended usage context changes?
    
    \item \textbf{RQ2 (Feasibility and Reliability):} Does feasibility validation reduce non-executable or unrealistic rule generation and improve execution stability?
    
    \item \textbf{RQ3 (Outcome Quality and Auditability):} Are the produced assessment results explainable, actionable, and traceable to executed validation rules?
    
    \item \textbf{RQ4 (Retrieval-Grounded Stability):} Does retrieval-augmented grounding constrain agentic generation such that assessment logic remains aligned with previously validated assessment semantics?
\end{itemize}

\subsection{Dataset}
\label{sec:eval_dataset}

The first three evaluation scenarios (RQ1, RQ2, RQ3) are conducted on the \textbf{German Credit dataset} \cite{german_credit}, a widely used benchmark in credit risk analysis. The dataset contains approximately 1,000 records and a mixture of numerical and categorical attributes, including \textit{Age}, \textit{Sex}, \textit{Job}, \textit{Housing}, \textit{Saving accounts}, \textit{Checking account}, \textit{Credit amount}, \textit{Duration}, and \textit{Purpose}.

The dataset exhibits realistic quality issues, including missing values in financial attributes (notably \textit{Saving accounts} and \textit{Checking account}) and categorical inconsistencies in \textit{Purpose}. No cleaning or preprocessing is applied prior to assessment in order to preserve native quality signals and evaluate the framework under realistic conditions.

\subsection{Usage Scenarios and User Interaction}
\label{sec:eval_scenarios}

To evaluate context sensitivity, the same dataset is assessed under three distinct usage scenarios. Each scenario is specified exclusively through a natural-language usage description provided by the user via the system interface.

\begin{figure}[!htbp]
    \centering
    \includegraphics[width=0.8\linewidth]{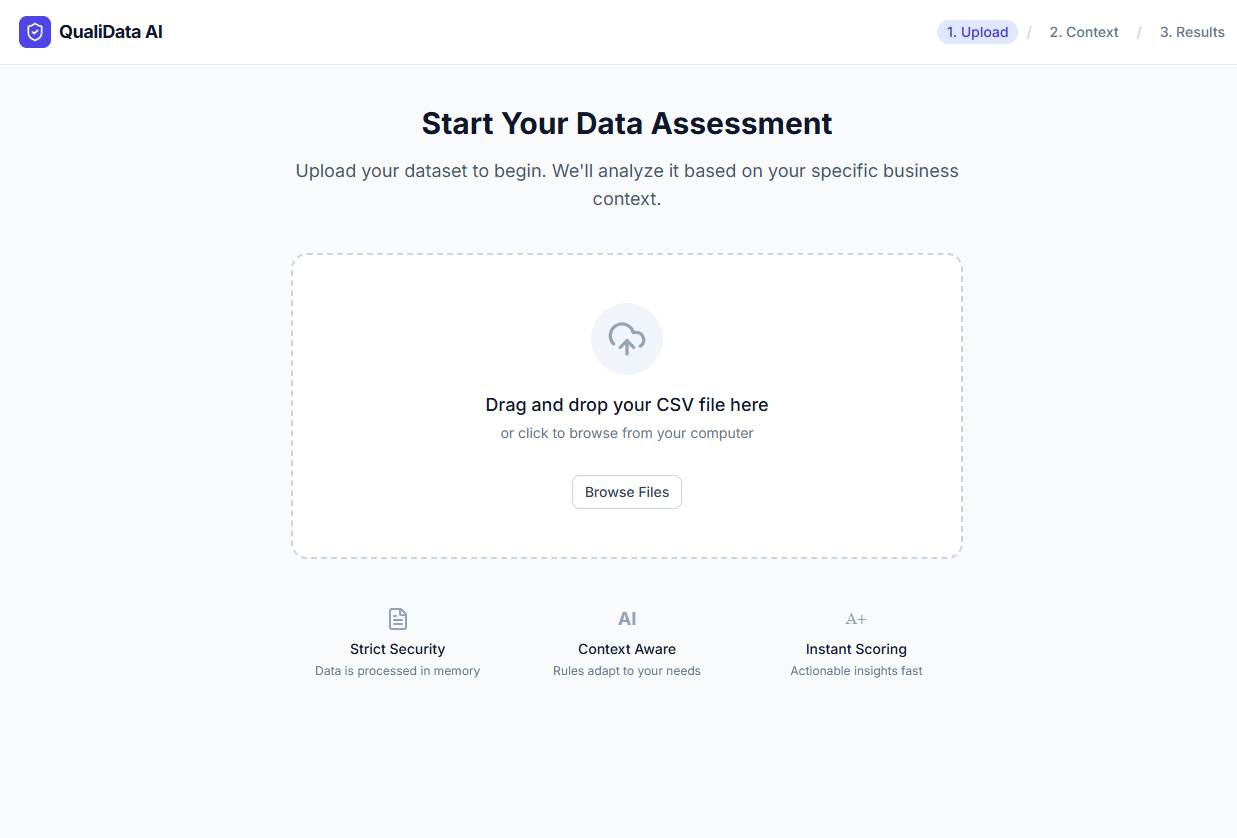}
    \caption{Dataset upload interface of the prototype.}
    \label{fig:upload_ui}
\end{figure}

\begin{figure}[!htbp]
    \centering
    \includegraphics[width=0.8\linewidth]{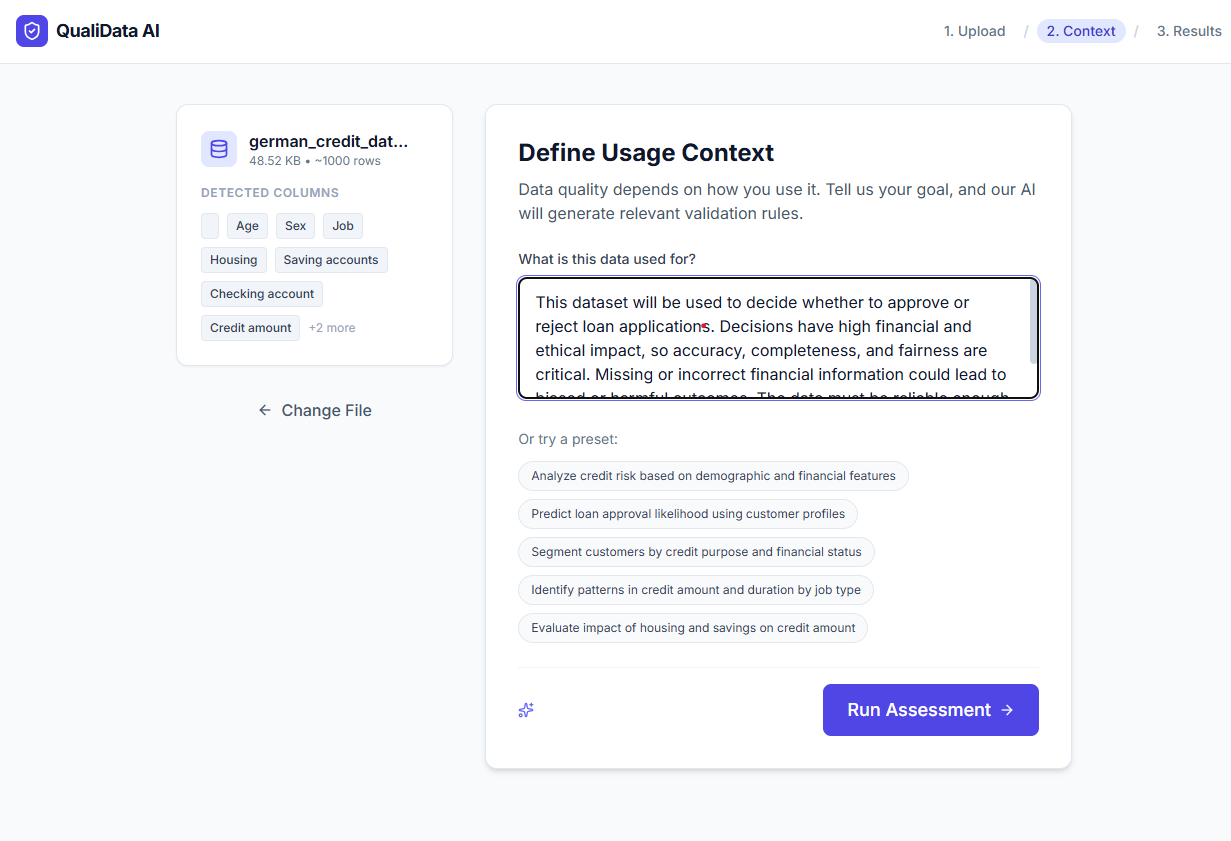}
    \caption{Natural-language usage specification for the business decision (loan approval) scenario.}
    \label{fig:context_business}
\end{figure}

\begin{figure}[!htbp]
    \centering
    \includegraphics[width=0.8\linewidth]{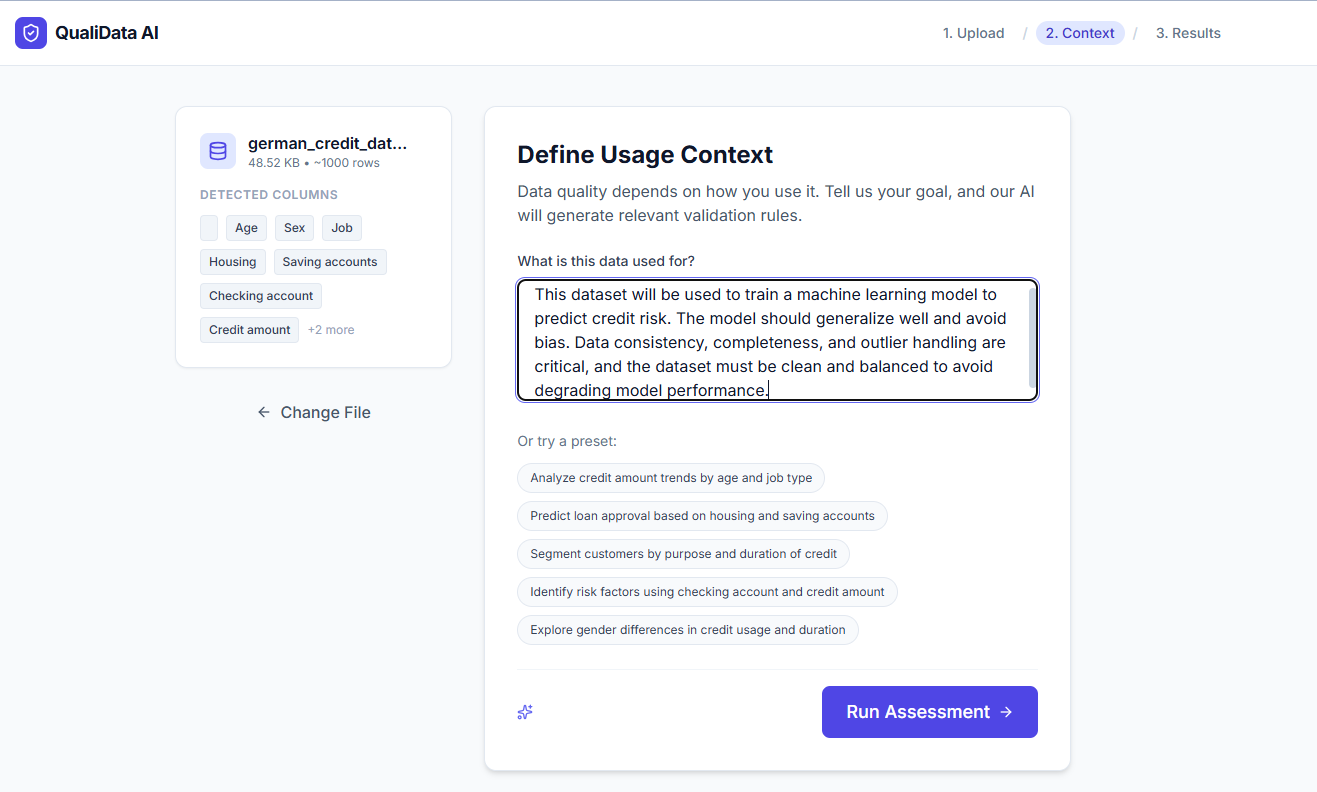}
    \caption{Natural-language usage specification for the machine learning scenario.}
    \label{fig:context_ml}
\end{figure}

Figure~\ref{fig:upload_ui} shows the dataset upload interface, while Figures~\ref{fig:context_business} and~\ref{fig:context_ml} illustrate how users provide detailed usage context that directly drives assessment planning.

The evaluated scenarios are:

\begin{itemize}
    \item \textbf{Business Decision Scenario}: the dataset is used to approve or reject loan applications, where financial risk, fairness, and ethical impact are critical.
    \item \textbf{Machine Learning Scenario}: the dataset is used to train predictive models for credit risk estimation, emphasizing consistency, completeness, and robustness.
    \item \textbf{Regulatory Auditing Scenario}: the dataset is used to assess fairness and non-discrimination in lending decisions, prioritizing demographic integrity and interpretability.
\end{itemize}

\subsection{Results I: Context-Sensitive Assessment Behavior (RQ1)}
\label{sec:eval_rq1}

The evaluation demonstrates that the framework produces substantially different assessment strategies and outcomes for the same dataset when the intended usage context changes.

\begin{figure}[!htbp]
    \centering
    \includegraphics[width=0.8\linewidth]{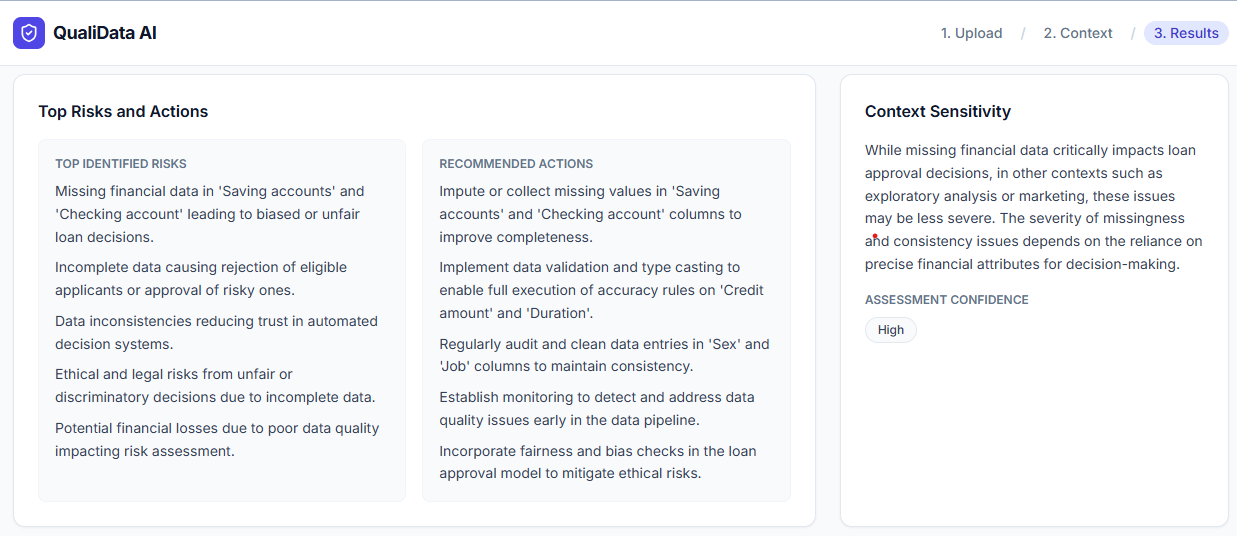}
    \caption{Top risks and recommended actions for the business decision scenario.}
    \label{fig:bd_risks}
\end{figure}

\begin{figure}[!htbp]
    \centering
    \includegraphics[width=0.8\linewidth]{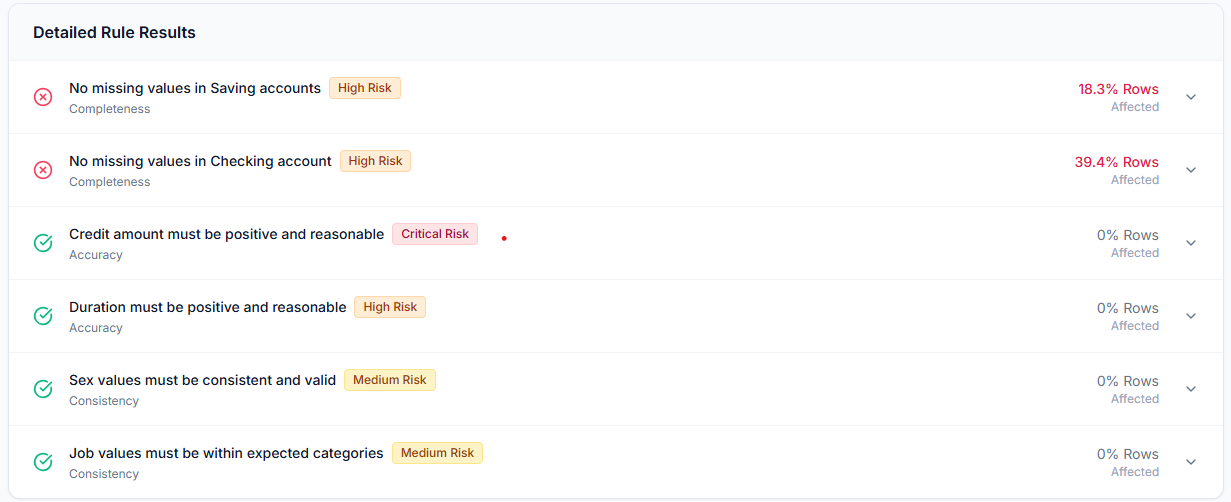}
    \caption{Rule-level validation results for the business decision scenario.}
    \label{fig:bd_rules}
\end{figure}

\begin{figure}[!htbp]
    \centering
    \includegraphics[width=0.8\linewidth]{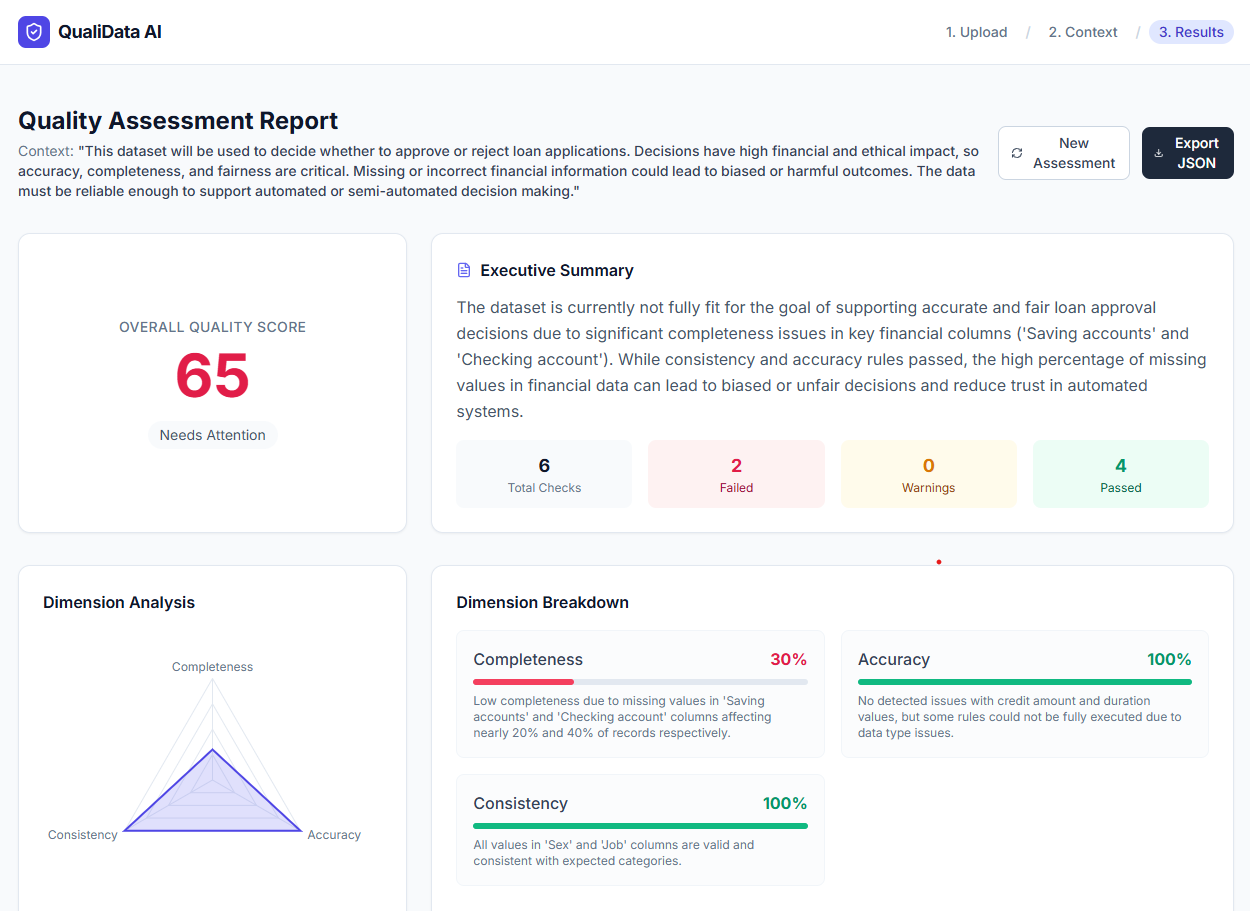}
    \caption{Overall assessment summary for the business decision scenario.}
    \label{fig:bd_summary}
\end{figure}

\paragraph{Business decision scenario:}
In the loan approval context, the system prioritizes completeness and accuracy of financial attributes. As shown in Figures~\ref{fig:bd_risks}--\ref{fig:bd_summary}, missing values in \textit{Saving accounts} (18.3\% of records) and \textit{Checking account} (39.4\%) are flagged as high-risk issues, leading to a reduced overall quality score (65) and a classification of \emph{Needs Attention}. The generated explanations emphasize ethical and financial risks associated with biased or unreliable decisions.

\begin{figure}[!htbp]
    \centering
    \includegraphics[width=0.8\linewidth]{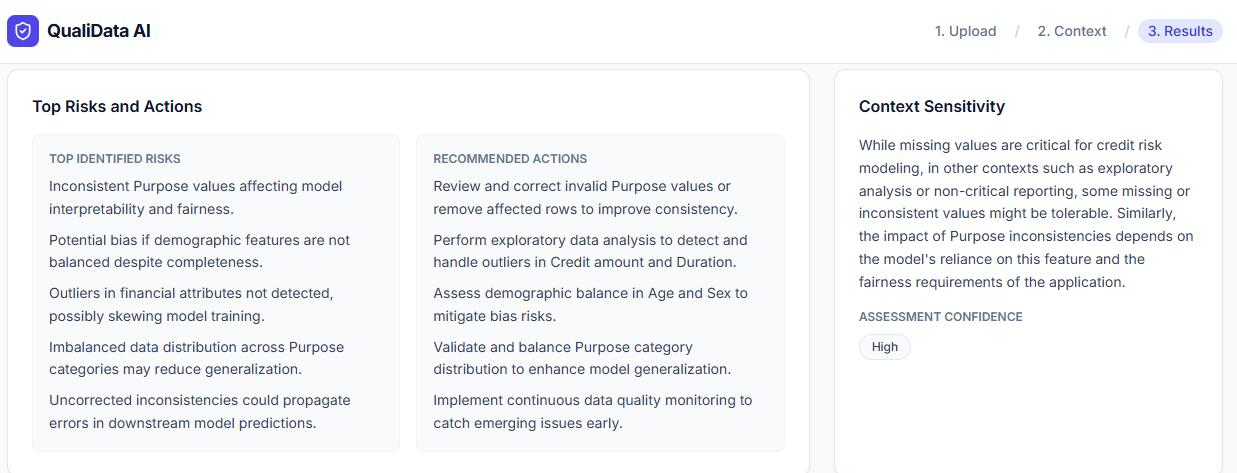}
    \caption{Top risks and actions for the machine learning scenario.}
    \label{fig:ml_risks}
\end{figure}

\begin{figure}[!htbp]
    \centering
    \includegraphics[width=0.8\linewidth]{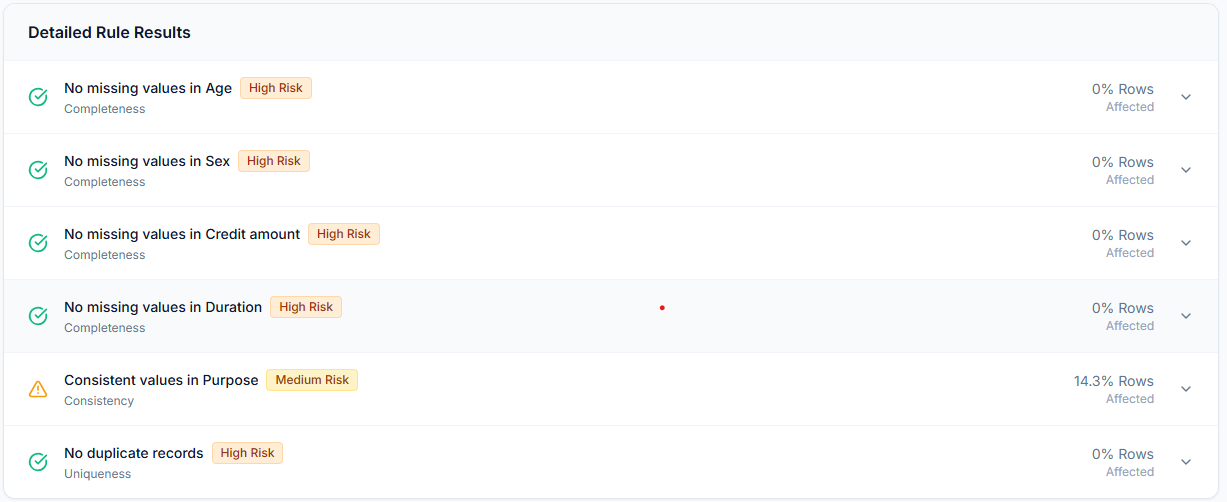}
    \caption{Primary rule-level results for the machine learning scenario.}
    \label{fig:ml_rules}
\end{figure}

\begin{figure}[!htbp]
    \centering
    \includegraphics[width=0.8\linewidth]{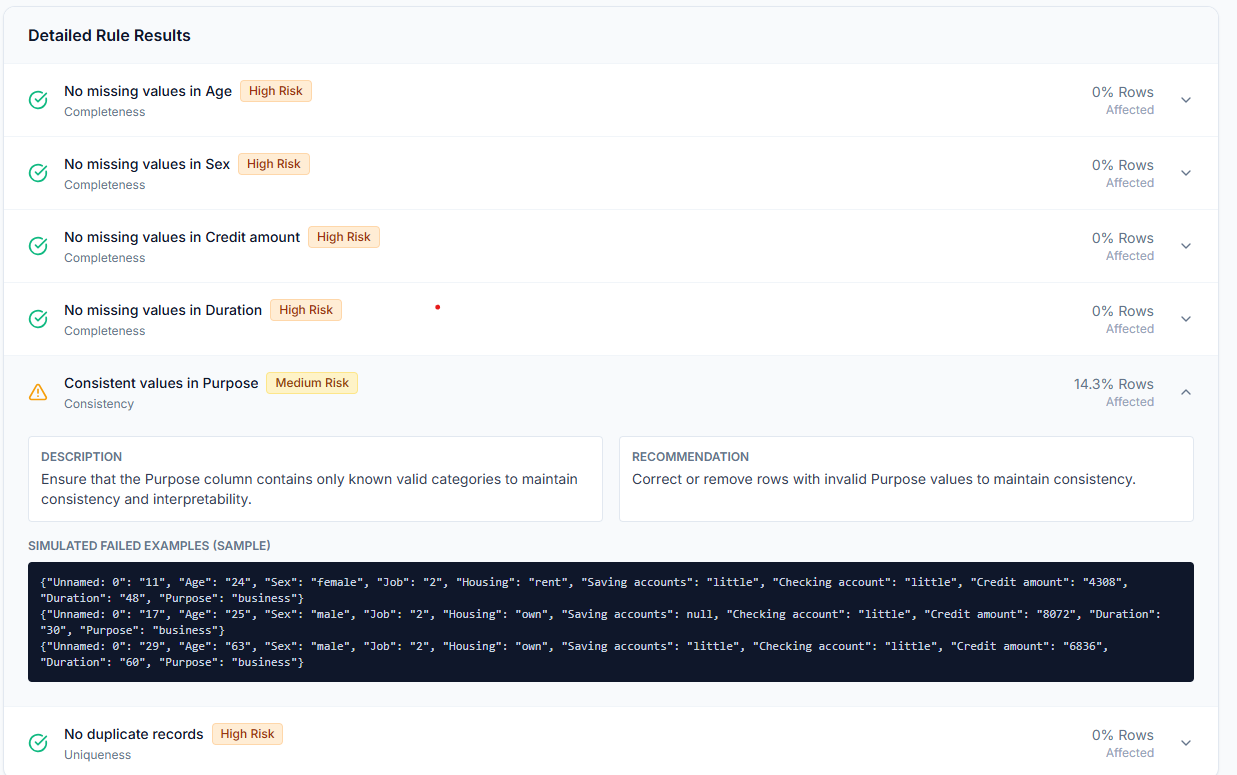}
    \caption{Detailed rule inspection for categorical consistency in the machine learning scenario.}
    \label{fig:ml_rules2}
\end{figure}

\begin{figure}[!htbp]
    \centering
    \includegraphics[width=0.8\linewidth]{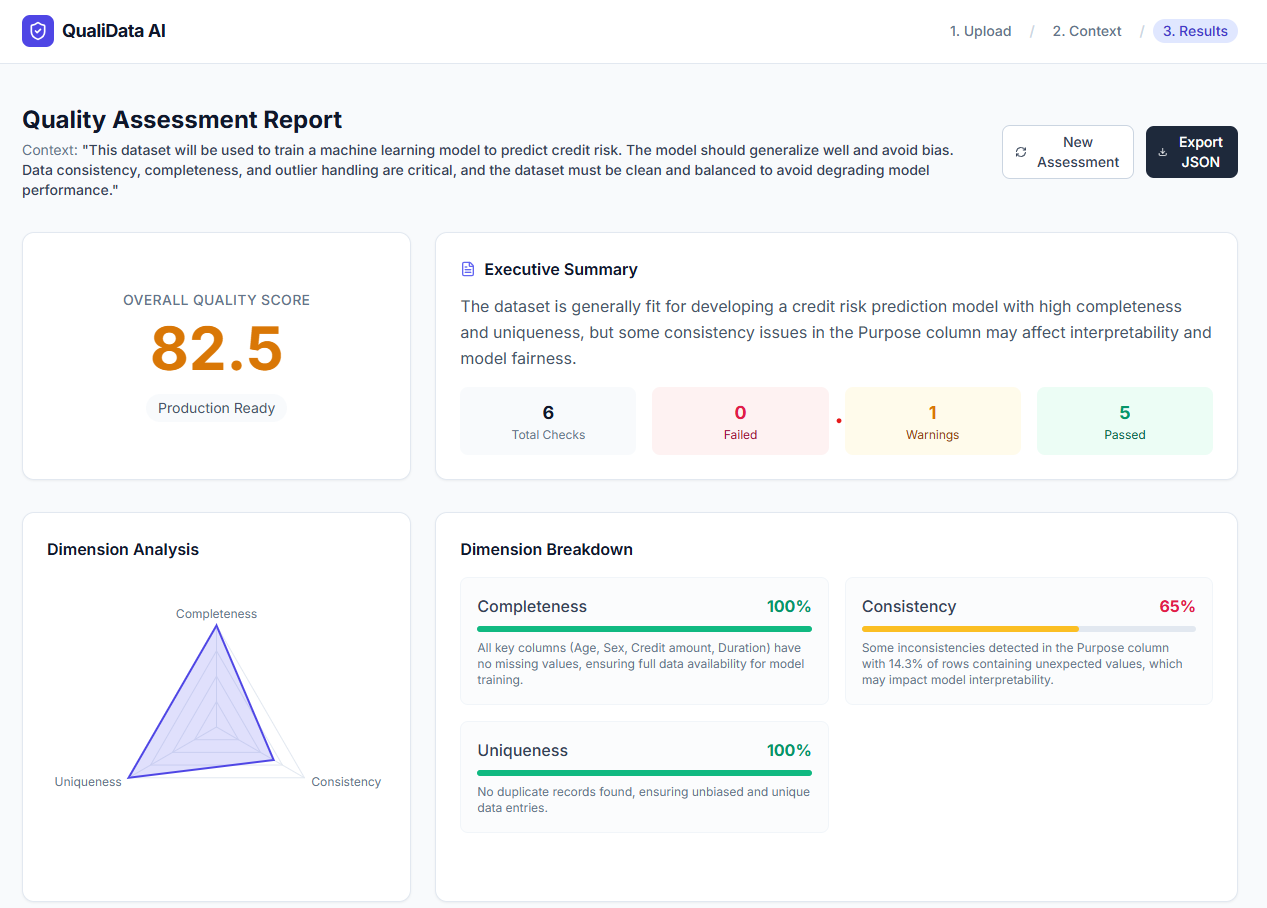}
    \caption{Overall assessment summary for the machine learning scenario.}
    \label{fig:ml_summary}
\end{figure}

\paragraph{Machine learning scenario:}
In the machine learning context, the assessment shifts toward consistency, uniqueness, and distributional stability. Figures~\ref{fig:ml_risks}--\ref{fig:ml_rules2}--\ref{fig:ml_summary} show that while most completeness checks pass, inconsistencies in the \textit{Purpose} attribute (affecting 14.3\% of records) are highlighted due to their impact on model interpretability and fairness. The overall quality score is higher than in the business decision scenario, reflecting suitability for model training with targeted preprocessing.

\begin{figure}[!htbp]
    \centering
    \includegraphics[width=0.8\linewidth]{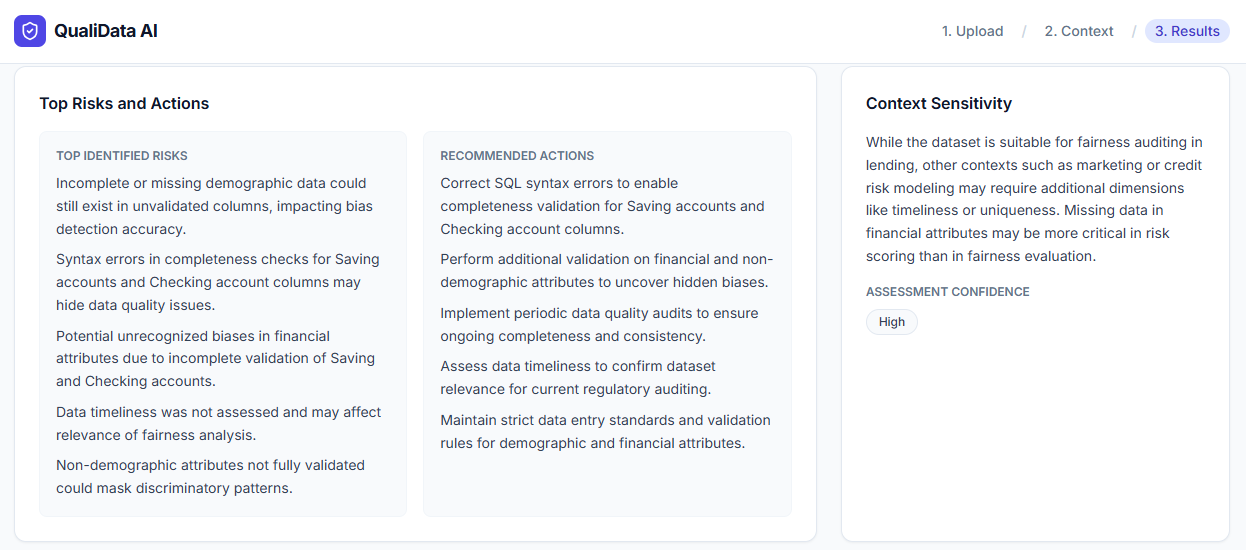}
    \caption{Identified risks and recommendations for the regulatory auditing scenario.}
    \label{fig:audit_risks}
\end{figure}

\begin{figure}[!htbp]
    \centering
    \includegraphics[width=0.8\linewidth]{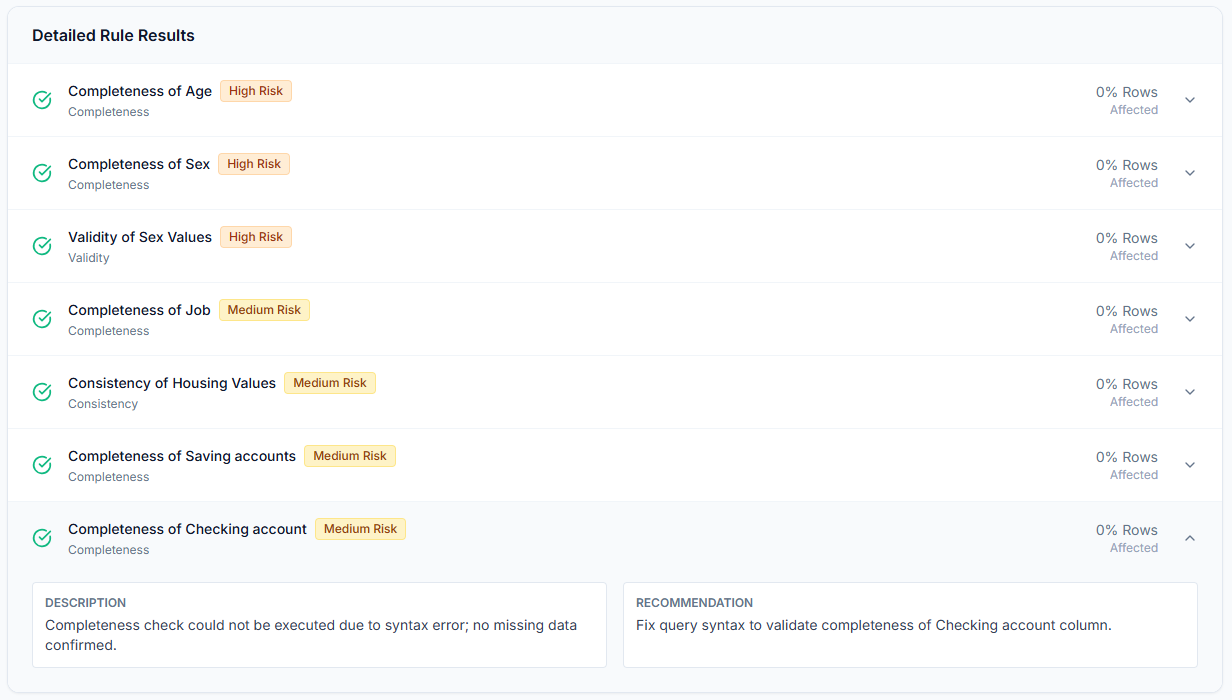}
    \caption{Rule-level validation outcomes for the auditing scenario.}
    \label{fig:audit_rules}
\end{figure}

\begin{figure}[!htbp]
    \centering
    \includegraphics[width=0.8\linewidth]{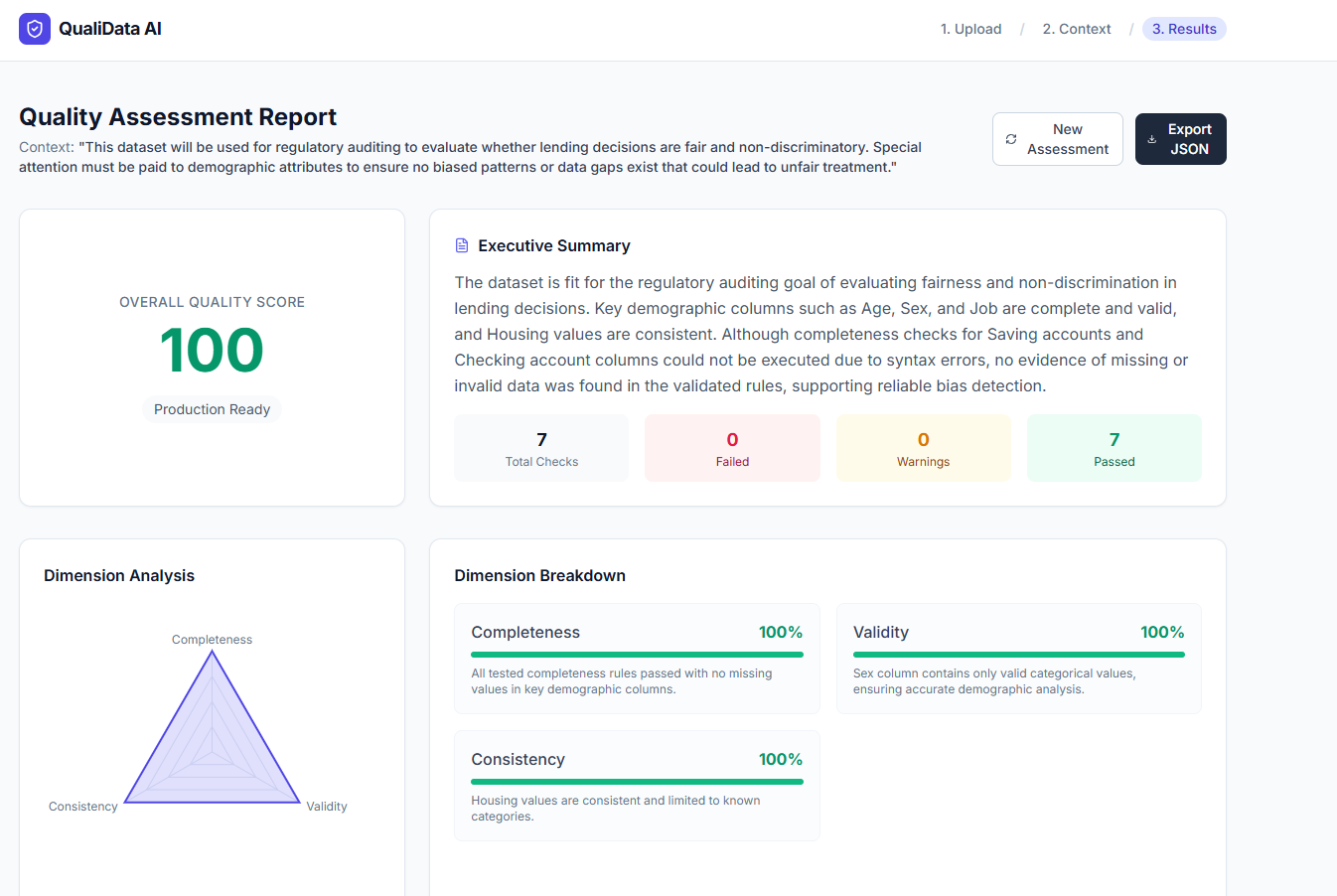}
    \caption{Overall assessment summary for the regulatory auditing scenario.}
    \label{fig:audit_summary}
\end{figure}

\paragraph{Regulatory auditing scenario:}
In the auditing context, the system prioritizes demographic completeness, validity, and categorical consistency. Figures~\ref{fig:audit_risks}--\ref{fig:audit_summary} show that demographic attributes (\textit{Age}, \textit{Sex}, \textit{Job}) are complete and valid, resulting in an overall quality score of 100 and a \emph{Production Ready} classification. Financial attributes with unexecuted checks are transparently reported without affecting the fairness assessment.

This demonstrates that assessment behavior is driven by intended usage rather than static dataset properties, directly addressing RQ1.

\subsection{Results II: Feasibility Validation (RQ2)}
\label{sec:eval_feasibility}

To evaluate the impact of feasibility validation on execution reliability (RQ2), we analyze the framework's behavior when encountering rules that, while logically sound, conflict with the observed characteristics of the dataset. The core strength of the Feasibility Validator Agent lies not just in filtering syntax errors, but in applying domain-specific data context to gate unrealistic constraints before they reach the deterministic execution layer.

A critical example of this capability was observed during the evaluation on the German Credit dataset, specifically regarding a proposed rule for the \textit{Credit amount} attribute.

\begin{figure}[!htbp]
    \centering
    \includegraphics[width=0.8\linewidth]{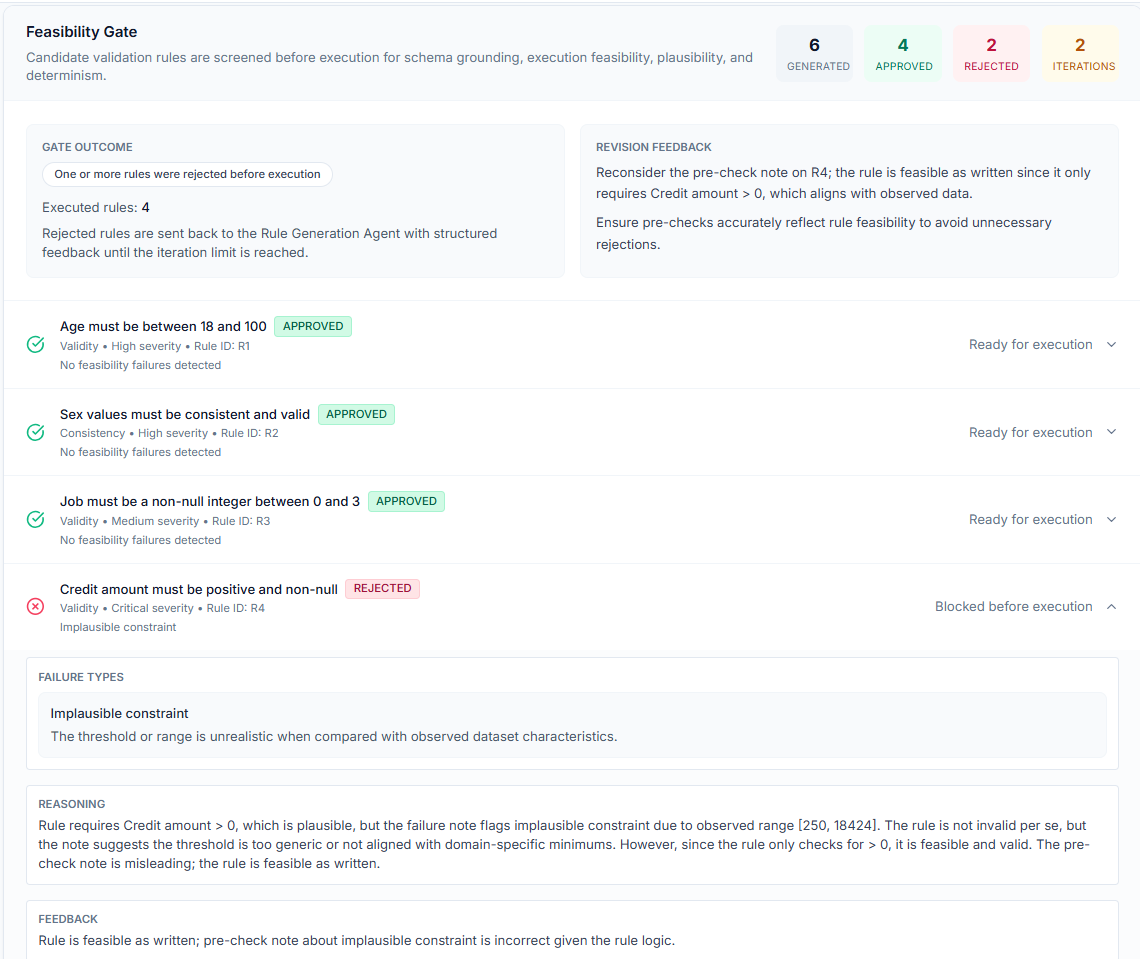}
    \caption{Full interface overview showing the Feasibility Gate status (6 generated, 4 approved, 2 rejected) and the detailed failure blocks.}
    \label{fig:feasibility_gate_all}
\end{figure}

\paragraph{Infeasible Rule Detection and Iterative Refinement:}
Figure~\ref{fig:feasibility_gate_all} provides a comprehensive view of the Feasibility Gate interface, which serves as the primary mechanism for mitigating the risk of executing non-viable logic. In this execution instance, the system processed four executable rules. While three rules were immediately approved, one rule—Rule ID R4: ``Credit amount must be positive and non-null''—was flagged as \textbf{REJECTED} by the Feasibility Validator.

The system's detailed analysis of this failure, excerpted in Figure~\ref{fig:feasibility_gate_all}, illustrates the multi-step reasoning of the validator agent:

\begin{itemize}
    \item \textbf{Failure Type:} The agent categorized the issue as an \textit{Implausible constraint}. The accompanying system feedback defines this as a scenario where ``The threshold or range is unrealistic when compared with observed dataset characteristics.''
    \item \textbf{Validator Reasoning:} The validator's internal reasoning note (visible in Figure~\ref{fig:feasibility_gate_all}) provides the crucial context. It acknowledges that the abstract rule logic (Credit amount $>0$) is plausible. However, it explicitly flags a conflict with the detected distribution of the data: \textit{flags implausible constraint due to observed range [250, 18424]}. 
    \item \textbf{Refinement Loop Feedback:} The reasoning concludes that because the actual data starts at 250, a check merely for $>0$ is ``too generic.'' This triggers specific \textit{Revision Feedback} (visible in the top-right of Figure~\ref{fig:feasibility_gate_all}) instructing the Rule Generation Agent to ``Ensure pre-checks accurately reflect rule feasibility to avoid unnecessary rejections.''
\end{itemize}

This interaction confirms that the Feasibility Validator successfully prevents the deployment of naive constraints by grounding the assessment logic in the statistical reality of the data.

\paragraph{Impact on Reliability:}
Without this gating mechanism, the system would have executed a redundant or misleading check. The observed behavior demonstrates that the feasibility validation stage (operating in Iteration 2 in this example) successfully identifies and halts logically valid but operationally unrealistic rules. 

By enforcing this contextual check prior to execution, the framework ensures that the final assessment reports are derived only from high-fidelity, validated specifications. This directly addresses RQ2 by proving that the Feasibility Gate significantly enhances the operational reliability and trustworthiness of the autonomous assessment workflow.

\subsection{Results III: Explainability and Auditability (RQ3)}
\label{sec:eval_rq3}

Across all scenarios, the framework produces structured assessment reports that include rule-level outcomes, affected-record percentages, dimension-level scores, overall quality scores, and actionable recommendations. All reported insights are derived exclusively from executed validation logic.

The consistency between rule outcomes, summarized risks, and recommendations demonstrates strong traceability. Users are able to inspect individual failed or warning rules, view example affected records, and understand how these outcomes influence the final quality score. This satisfies RQ3 by demonstrating explainable, auditable, and reproducible assessment outcomes.

\subsection{Results IV: Retrieval-Augmented Grounding (RQ4)}
\label{sec:eval_rag}

To evaluate the effect of retrieval-augmented grounding on assessment stability and semantic alignment, we conduct an additional experiment using a \emph{healthcare patient records dataset}. The objective is to assess whether the retrieval component constrains agentic generation such that executable validation logic remains faithful to previously validated assessment semantics, particularly in a safety-critical domain.

\paragraph{Healthcare dataset:}
The evaluated dataset consists of a small tabular extract of patient records (100 rows, 12 attributes), including patient identifiers, demographic attributes, medical device identifiers, clinical observations, laboratory results, and adverse event descriptions. The dataset intentionally exhibits realistic quality challenges, including missing values in clinical attributes, mixed data representations (e.g., formatted vital signs), and identifier consistency constraints.

\paragraph{Retrieval-augmented prompt construction:}
At runtime, a lightweight context representation is extracted from the input dataset, capturing structural metadata (row and column counts), attribute names, data types, and missingness indicators. This context is embedded and used to retrieve the most semantically similar healthcare assessment context ($k=1$) from the Neo4j-backed repository via the Graphiti retrieval layer.

\begin{figure}[!htbp]
\centering
\includegraphics[width=\columnwidth]{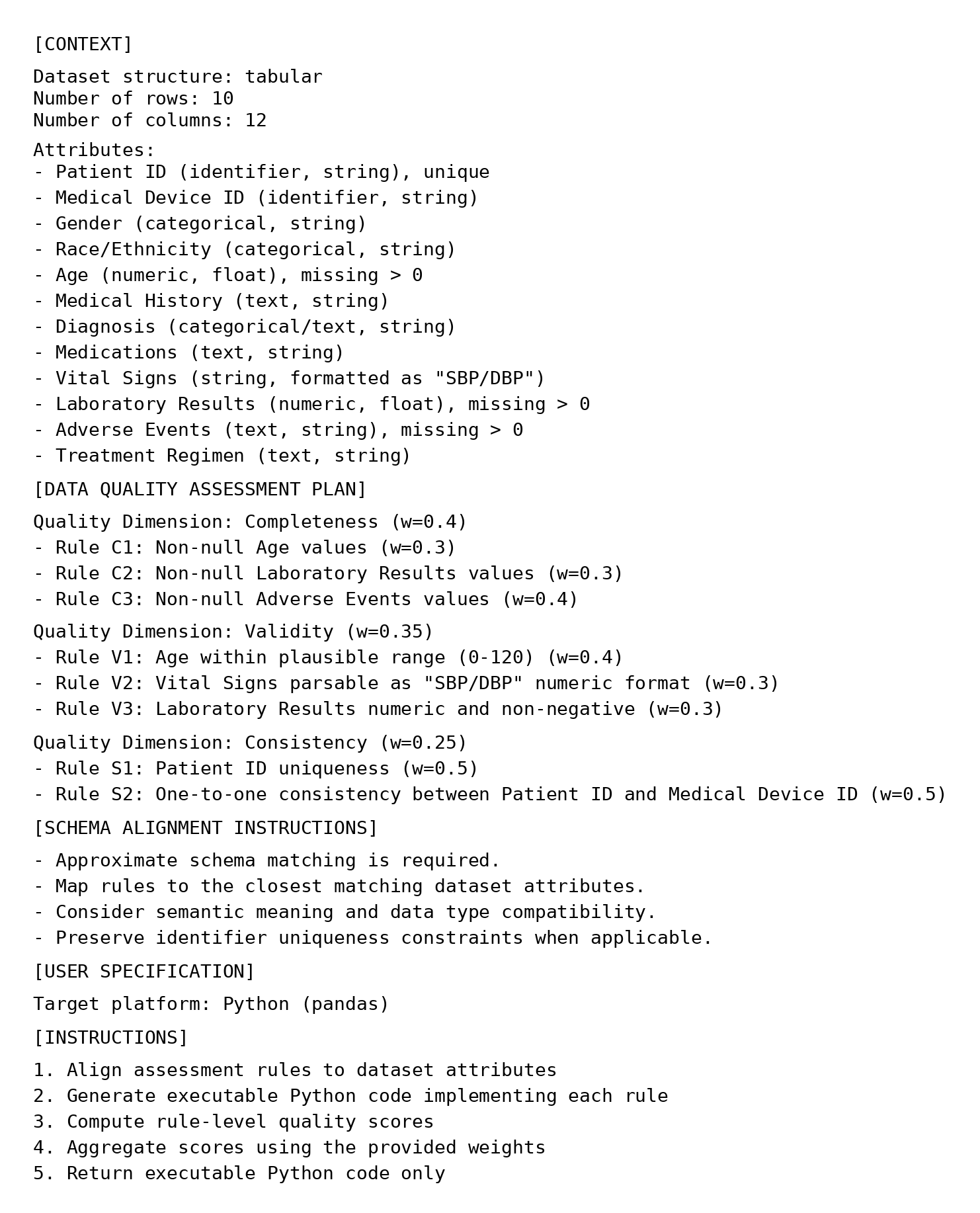}
\caption{Example of an augmented prompt constructed for a healthcare patient records dataset, illustrating context-sensitive grounding of executable validation logic under the proposed framework.}
\label{lst:augmented_prompt_healthcare}
\end{figure}

The retrieved assessment plan—comprising predefined quality dimensions, rule semantics, and weighting schemes—is then combined with the extracted dataset context to form an augmented prompt. Figure~\ref{lst:augmented_prompt_healthcare} shows the resulting prompt used to constrain the language model.

The prompt explicitly specifies completeness, validity, and consistency requirements relevant to patient safety and clinical integrity, while enforcing approximate schema alignment and preservation of identifier constraints.

\paragraph{Generated execution artifact:}
Given the augmented prompt, the language model generates executable Python (pandas) validation code implementing the retrieved assessment plan. Inspection of the generated artifact confirms that:
(i) all rules correspond directly to retrieved plan components,
(ii) no additional quality dimensions or unsupported checks are introduced, and
(iii) all constraints are translated into deterministic, row-level validation logic.

The generated code correctly implements completeness checks for clinical attributes (e.g., Age, Laboratory Results, Adverse Events), validity checks for physiological ranges and formatted vital signs, and consistency checks enforcing patient identifier uniqueness and one-to-one device associations. Rule-level scores, dimension-level aggregation, and overall quality computation strictly follow the retrieved weighting scheme.

\paragraph{Observations:}
The healthcare experiment demonstrates that retrieval-augmented grounding effectively stabilizes agentic generation in a sensitive domain. Despite the expressive freedom of the language model, the produced validation logic remains tightly aligned with retrieved assessment semantics and exhibits no hallucinated rules or dimensions. This confirms that the RAG component serves as an authoritative constraint rather than a generative suggestion mechanism.

Overall, these results provide empirical support for \textbf{RQ4}, showing that retrieval-augmented grounding enables context-sensitive yet semantically controlled execution, even in domains where correctness, traceability, and safety are critical.

\subsection{Discussion}

Overall, the evaluation demonstrates that the proposed unified agentic--retrieval framework produces context-sensitive, explainable, and reliable data quality assessments. The results demonstrate that controlled agentic reasoning, when combined with retrieval grounding and feasibility validation, enables reliable and context-sensitive data quality assessment in practical settings.
Interestingly, as seen in the \textit{Revision Feedback} (Figure~\ref{fig:feasibility_gate_all}), the system identified a nuance where the validator was perhaps \textit{over-constrained}. While the observed range [250, 18424] makes a ``$>0$'' rule technically valid, the agent flagged it as too generic. This highlights a critical trade-off in autonomous data management: the balance between strict statistical grounding and logical permissiveness. In a production environment, this self-correction mechanism prevents the generation of ``trivial'' rules that satisfy basic logic but fail to capture meaningful data anomalies.

\section{Conclusion and Future Work}
\label{sec:conclusion}

Data quality assessment is a critical prerequisite for reliable analytics and data-driven decision-making, yet it remains highly context-dependent and difficult to operationalize in an automated and reproducible manner.

This paper proposed a unified agentic--retrieval framework for autonomous context-aware data quality assessment. The proposed framework operationalizes natural-language descriptions of intended data usage into structured assessment intent, executable validation logic, and interpretable results through a multi-agent workflow. A key contribution is the feasibility validation gate, which enforces execution-aware filtering of generated specifications by identifying and rejecting infeasible or unrealistic rules prior to execution. To preserve auditability and reproducibility, the framework strictly separates probabilistic reasoning from deterministic execution and bases final explanations exclusively on executed outcomes.

The evaluation demonstrates that the framework adapts assessment behavior to different usage intents and that feasibility-gated execution effectively prevents unrealistic rule generation and execution failures. These results support the practicality of combining agentic orchestration with retrieval-augmented grounding and controlled execution to bridge the gap between conceptual context-aware assessment planning and reliable operational validation.

Future work will focus on two concrete directions. First, the framework will be evaluated across multiple domains and datasets to study generality, robustness, and transferability of retrieved assessment knowledge beyond a single dataset or application setting. Second, we will strengthen validation and refinement loops by incorporating systematic self-checking and verification mechanisms. Furthermore, we plan to extend the deterministic execution layer to support alternative engines—such as Apache Spark or Dask—to enable big data scalability and handle large-scale distributed datasets that exceed local memory constraints.

\section{Declarations}

\subsection{Conflict of Interest}
All authors declare that they have no conflicts of interest.

\subsection{Use of AI Technology}
During the preparation of this manuscript, the authors used AI-based tools, namely ChatGPT and Gemini, for language editing tasks such as correcting grammatical errors, improving clarity, and minimizing redundancy. These tools were not used to generate original research content, perform analysis, or draw conclusions. 

\bibliographystyle{unsrt}
\bibliography{Ref}

\end{document}